\documentclass[aps,prl,twocolumn, showpacs,superscriptaddress]{revtex4-2}
\usepackage{amsmath}
\usepackage{xcolor}
\usepackage{graphicx}
\usepackage{dcolumn}
\usepackage{gensymb}
\usepackage{float}
\usepackage{bm}
\usepackage{hyperref}
\begin{document}

\title{Dissipationless Circulating Currents and  Fringe Magnetic Fields Near a Single Spin Embedded in a  Two-Dimensional Electron Gas} 

\author{Adonai R. da Cruz}
 \email{adonairc@gmail.com}
\affiliation{Department of Applied Physics, Eindhoven University of Technology, Eindhoven 5612 AZ, The Netherlands}

\author{Michael E. Flatt\'{e}}
\email{michael\_flatte@mailaps.org}
\affiliation{Department of Physics and Astronomy,  University of Iowa, Iowa City, Iowa 52242, USA}
\affiliation{Department of Applied Physics, Eindhoven University of Technology, Eindhoven 5612 AZ, The Netherlands}

\date{\today}

\begin{abstract}
Theoretical calculations predict the dissipationless circulating current induced by a spin defect in a two-dimensional electron gas with 
spin-orbit coupling. The shape and spatial extent of these dissipationless circulating currents depends dramatically on the relative strengths of  spin-orbit fields with differing spatial symmetry, offering the potential to use an electric gate to manipulate nanoscale magnetic fields and couple  magnetic defects. The spatial structure of the magnetic field produced by this current is calculated and provides a direct way to measure the  spin-orbit fields of the host, as well as the defect spin orientation, \textit{e.g.} through scanning nanoscale magnetometry.

\end{abstract}

\maketitle
Single spins associated with point defects in solid-state materials are promising candidates for  qubits and novel quantum spintronic devices for communication, sensing, and information processing \cite{Kane1998AComputer,Loss1998,Awschalom2002,Ichimura2001AMode,Wolfowicz2021QuantumDefects}. Isolated magnetic dopants embedded in two-dimensional electron gases (2DEG) with spin-orbit coupling (SOC) can be electrically and optically addressed \cite{Matsui2007,Lee2010,Myers2008Zero-fieldSemiconductors}  and coherently manipulated. Control of the relative phases of defect spin wave functions through manipulation of the spin-spin coupling enables spin-based quantum computation\cite{Loss1998,Kane1998AComputer,Awschalom2002}. To date most spin-spin couplings are achieved either through dipolar magnetic fields arising from the spin \cite{Gaebel2006,Fuchs2011}, which are difficult to tune, or through photonic coupling \cite{Togan2010}, which is effective for much longer distances. The magnetic moment of a bare spin is modified by the spin-orbit interaction of the host, leading to proposals for single-spin control via $g$ tensor manipulation \cite{Kato2003,Pingenot2008,Pingenot2011,Schroer2011,Takahashi2013,Ares2013,Crippa2018,Ferron2019}, however the effects on 
nanoscale dipolar fields and the consequences for coupling spins remain unexplored.

Here we derive the dissipationless circulating current surrounding a spin embedded in a 2DEG with spin-orbit coupling, and show that these effects are significant and their spatial structure is detectable by current  nanoscale 
magnetometry\cite{Doherty2013TheDiamond,Grinolds2013NanoscaleConditions,Casola2018ProbingDiamond,Degen2017}.  The spin-orbit interaction in a 2DEG can be modified with a perpendicular electric field (without  dissipative current flow) thus we further propose that modifying these magnetic fields provides a method of electrically tuning spin-spin interactions to produce quantum entangling gates. Our results rely on a derivation of the 2DEG's current operator  that identifies  the critical role of effective spin-orbit vector potentials in correctly producing currents that are dissipationless. We compare diverse hosts for  these phenomena including semiconductor quantum wells (QWs) and surfaces (e.g., the electron accumulation layer in InAs \cite{Noguchi1991InMinato,Bell1997AccumulationSurfaces,Canali1998Low-temperatureSurfaces}), 
oxide interfaces (e.g., LaAlO\textsubscript{3}/SrTiO\textsubscript{3}\cite{Fete2014LargeInterface}), atomically thin materials (e.g., BiSb\cite{Singh_BiSb} and  WSeTe\cite{Yao2017ManipulationDichalcogenides})
and films (e.g., LaOBiS\textsubscript{2}\cite{Liu2013TunableTransistors}).  
Our  expressions for   circulating currents connect  to other fundamental features of an electron gas interacting with localized magnetic moments, including  Friedel oscillations of  charge and spin densities \cite{Friedel1958MetallicAlloys,Lounis2012MagneticWaves}, the Dzyaloshinskii-Moriya interaction \cite{Freimuth2017} 
and long-range Ruderman Kittel-Kasuya-Yosida \cite{Ruderman1954IndirectElectrons,Kasuy1956AModel,Yosida1957MagneticAlloys,Chesi2010RKKYCouplings} interactions between defects. Measurement of the spatial structure of this current can be achieved by probing the magnetic fringe field above the surface with nanoscale scanning magnetometry (e.g. using nitrogen-vacancy (NV) centers in diamond \cite{Doherty2013TheDiamond,Grinolds2013NanoscaleConditions,Casola2018ProbingDiamond,Degen2017}), which have achieved spatial resolution in the nanometer range and field sensitivity (tens to hundreds of nT), 
on the order of the  features calculated here. Another option relies on recent advances in electron ptychography for direct sensing\cite{Chen2021}.

Central to our results is the derivation of the current operator associated with the effective Hamiltonian describing a 2DEG with SOC,
\begin{equation}
\label{eq:ham}
H=\frac{\hbar^{2}k^{2}}{2m^{*}}+\alpha\left(\sigma_{x}k_{y}-\sigma_{y}k_{x}\right)+\beta\left(\sigma_{x}k_{x}-\sigma_{y}k_{y}\right).
\end{equation}
 The  spin orbit fields linear in crystal momentum ${\bf k}$ emerge from both the crystal's bulk inversion asymmetry (BIA, with coefficient $\beta$)\cite{Dresselhaus1955Spin-OrbitStructures}  and the inversion asymmetry of the heterostructure (SIA, with coefficient $\alpha$, which can be tuned by applied electric fields perpendicular to the 2DEG plane)\cite{Bychkov1984OscillatoryLayers}.
 We derive the current operator (see Supplementary Material \cite{supp}) 
from the steady-state continuity equation  $\partial_{t}n + \nabla \cdot \mathbf{j} = 0$ and group the spin-orbit terms  together in an effective spin-dependent vector potential $\mathbf{A}_{so}$. 
The expected value of the charge current is evaluated using the Green's function (GF) scattering formalism,
\begin{widetext}
\begin{equation}
    \label{eq:current}
    \mathbf{j}\left(\mathbf{r}\right)=-\frac{e\hbar}{2\pi m^{*}}\text{Im}\int_{E_{so}}^{E_{f}}\left(t_{\uparrow}-t_{\downarrow}\right)\left\{ i\left[\left(\nabla G_{\uparrow\downarrow}^{0}\right)G_{\downarrow\uparrow}^{0}-\left(\nabla G_{\downarrow\uparrow}^{0}\right)G_{\uparrow\downarrow}^{0}\right]-\frac{2e}{\hbar}\left(\mathbf{A}_{so}^{\uparrow\downarrow}G_{\downarrow\uparrow}^{0}-\mathbf{A}_{so}^{\downarrow\uparrow}G_{\uparrow\downarrow}^{0}\right)G_{\uparrow\uparrow}^{0}\right\} dE.
\end{equation}
\end{widetext}
where $G^{0}\left(\mathbf{r},\mathbf{r}^{\prime}\right)$ are the 2DEG retarded GFs, $t_\sigma$ are the defect T-matrix elements with spin projection $\sigma$, $E_{so}$ is the energy minimum of the 2DEG, $E_f$ is the Fermi energy (determined by the electron density $n$)
and the effective vector potential
\begin{align}
    \mathbf{A}_{so} &=\frac{\hbar k_{so}}{e}
    \left\{\left[\sigma_{x}\cos\left(\theta-\tau\right)-\sigma_{y}\sin\left(\theta+\tau\right)\right]\hat{r} \right.\nonumber\\
    & \left. +\left[\sigma_{x}\sin\left(\theta-\tau\right)-\sigma_{y}\cos\left(\theta+\tau\right)\right]\hat{\theta} \right\},
\end{align}
with $\tau = \tan\left(\alpha/\beta\right)$ and $k_{so} = (m^{*}/\hbar^2)\sqrt{\alpha^2 + \beta^2}$.

The retarded Green's function from Eq.~(\ref{eq:ham}) is \cite{Berman2010Electron-beamWells}
\begin{equation}
\label{eq:gf_in_k}
g^{0}\left( \mathbf{k};E\right)=\left(\frac{2m^{*}}{\hbar^2}\right)\frac{(k^{2}_{E}-k^{2})\sigma_{0}-2k_{so}k[U_{x}\sigma_{x} - U_{y}\sigma_{y}]}{k^{4}_{E}-2k^{2}[k^{2}_{E}+2 k^{2}_{so}f_{\tau}(\theta_k)]+k^{4}}
\end{equation}
where $k^{2}_{E}=2m^{*}E/\hbar^{2}$  and $f_{\tau}\left(\theta_k\right)=1+\sin\left(2\tau\right)\sin\left(2\theta_{k}\right)$. The poles of the Green's function correspond to Fermi contours of the two spin-split subbands at energy $E$ and is obtained from the roots of the denominator,
\begin{equation}
k_{\pm} = Q \pm k_{so}\sqrt{f_{\tau}\left(\theta_k\right)},\  Q =\sqrt{k^{2}_{E}+k^{2}_{so}f_{\tau}\left(\theta_k\right)}.
\end{equation}

The  representation of the GF in real space is obtained by Fourier transform of Eq.~(\ref{eq:gf_in_k}),
\begin{widetext}
\begin{equation}
\label{eq:g0}
G^{0}\left(\mathbf{r};E\right)=-\frac{m^{*}}{2\pi^{2}\hbar^{2}}\int_{\theta-\pi/2}^{\theta+\pi/2}d\theta_{k}\left\{ \frac{\sigma_{0}}{Q}\left[k_{+}I_{1}\left(k_{+}\rho\right)+k_{-}I_{1}\left(k_{-}\rho\right)\right] -i\frac{\left[U_{x}\sigma_{x}-U_{y}\sigma_{y}\right]}{Q\sqrt{f_{\tau}}}\left[k_{+}I_{2}\left(k_{+}\rho\right)-k_{-}I_{2}\left(k_{-}\rho\right)\right]\right\},
\end{equation}
\end{widetext}
where $\rho = r\cos(\theta_k - \theta)$, $U_{x} =\cos(\theta_k - \tau)$, $U_{y} =\sin\left(\theta_k + \tau \right)$ and $I_{1/2}\left(k_{\pm}\rho\right)$ are closed-form functions \cite{Berman2010Electron-beamWells}.
Changes in the convexity of the outer contour, $k_{+}\left(\theta_{k},\tau\right)$, for a given energy $E=\sqrt{\hbar^{2}k_{E}^{2}}/2m^{*}$ are driven by the SOC ratio $\tau$ through $f_{\tau}\left(\theta_k\right)$ and
produce important consequences for electronic properties in real space, such as  enhanced electron beam formation for transport along symmetry axes\cite{Berman2010Electron-beamWells}.

We now embed a defect with spin employing the Dyson equation for the new GF perturbed by a magnetic potential. For simplicity we assume that the electron is scattered isotropically  with an incident wavelength large enough to not sense the spatial structure of the potential ($s$-wave approximation), although our treatment is easily generalized to more complex situations. The T-matrix\cite{Fiete2003Colloquium:Mirages}

\begin{equation}
\ t_\sigma = \left(\frac{4i\hbar^2}{m^{*}}\right)\left[e^{(2i\delta_{\sigma})} - 1\right],
\end{equation}
where the scattering phase-shift of the spin channel $\sigma$, $\delta_{\sigma}$, is related to the defect local density of states  through the Friedel sum rule \cite{Friedel1958MetallicAlloys} and encodes its charge and spin state.
Here we assume an energy independent T-matrix for a spin-polarized defect with magnetization fixed along the out-of-plane axis ($z$ axis), where the the spin-down state is fully occupied and does not participate in scattering processes ($\delta_\downarrow = 0$) and the partially occupied spin-up channel has a resonant level at the Fermi energy ($\delta_\uparrow = \pi/2$).  

We now show results for a 2DEG formed by the electron accumulation layer at a InAs(001) surface \cite{Noguchi1991InMinato}  using $k_{so} = 0.2\ \mathrm{nm}^{-1}$, $E_f = 10\ \mathrm{meV}$ and $m^{*} = 0.023\ m_{0}$. The low effective mass, strong spin-orbit coupling, and high Landé $g$ factor makes this a promising system in which to detect these  orbital features since the current magnitude is proportional to $({\alpha^2 + \beta^2})^{1/2}/m^{*}$ [Eq.~(\ref{eq:current})]. Conversion factors for different materials are provided in Table \ref{tab:table1} for the spatial dimensions ($d_{fac}$) and for the magnetic field strengths ($B_{fac}$) in the figures  below.

\begin{figure*}[hbt!]
\includegraphics[width=1\textwidth]{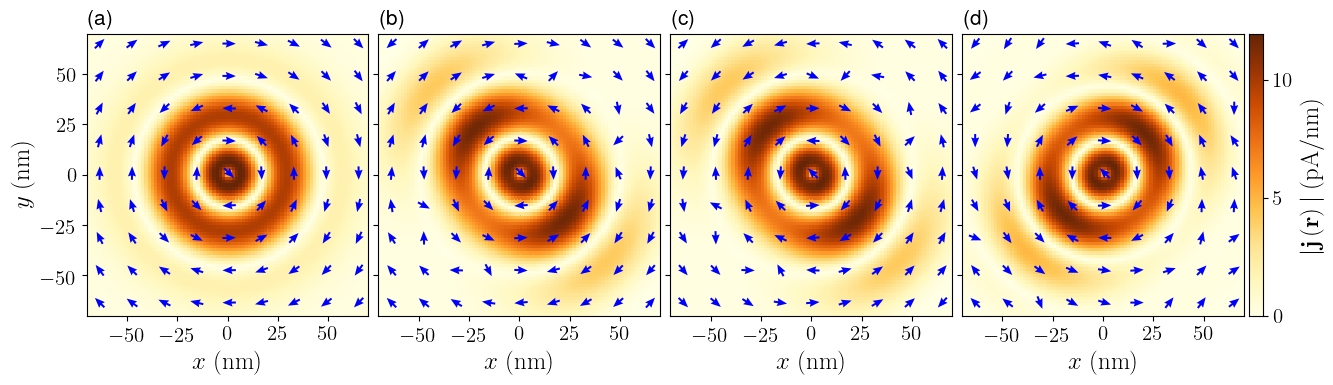}
\caption{Current density induced by a magnetic point defect with spin pointing perpendicular to an InAs 2DEG. (a) SIA dominated regime ($\beta=0$, $\tau = \pi/2$) showing angular symmetry.  (b-d) $\beta\ne 0$; the angularly anistropic circulating current stretches along the symmetry axis $\theta = 3\pi/4$ for $\alpha/\beta > 0$ (bc) or  along $\theta = \pi/4$ for $\alpha/\beta < 0$  (d). Opposite circulation direction of $\tau = \pi/4+0.1$ (b) from $\tau = \pi/4-0.1$ (c); current vanishes for  $\tau = \pi/4$ ($\alpha = \beta$). }
\label{fig:1}
\end{figure*}

 Figure~\ref{fig:1}  shows the current density induced by a magnetic defect at the origin ($\mathbf{r}_{0} = 0$)  with spin projection along the [001] out-of-plane direction, calculated from Eq.~(\ref{eq:current}). For $\alpha\gg \beta$,  corresponding to $\tau = \pi/2$ [Fig.~\ref{fig:1}(a)],
the isotropic dispersion relation implied by a $\theta_{k}$-independent $f_{\tau}$ allows for an  analytical solution to Eq.~(\ref{eq:g0}) in terms of Hankel functions of the first kind~\cite{Lounis2012MagneticWaves,Bouaziz2018Impurity-inducedGas}. The current then has only an angular component and oscillates with two characteristic lengths, reminiscent of Friedel oscillations, given by  $\lambda_{f}= \pi/k_{f}$ and $\lambda_{so} = \pi/k_{so}$.

The emergence of the anisotropy in the dispersion relation when both $\alpha$ and $\beta$ are relevant [Fig.~\ref{fig:1}(bc)] drastically changes the spatial structure  of the circulating current compared to  Fig.~\ref{fig:1}(a), stretching the features along the symmetry axes. For $\alpha/\beta > 0$ a stronger current density focuses along the $\theta = 3\pi/4$ direction  [Fig.~\ref{fig:1}(b)] and rotates by 90$^{\circ}$ focusing along $\theta = \pi/4$ when $\alpha/\beta < 0$ [Fig.~\ref{fig:1}(c)], a consequence of an interfering contribution of stationary points\cite{Berman2010Electron-beamWells}.
The  currents are equal but rotate in opposite directions for values of $\tau = \pi/4 \pm \delta$ ($\tau=\pi/4$ corresponds to $\alpha=\beta$), as shown in Fig.~\ref{fig:1}(bc). This change in rotation  would invert the  orbital magnetization around the spin (see~Supplementary Material\cite{supp}). We note that  when $\alpha = \beta$, the extra SU(2) symmetry implies a fixed spin quantization axis independent of $\mathbf{k}$, making the 2DEG GF of Eq.~(\ref{eq:g0}) spin-diagonal upon a global spin rotation \cite{Bernevig2006ExactSystem} and thus both contributions to the current in Eq.~(\ref{eq:current}) vanish.

\begin{figure}[hbt!]
  \includegraphics[width=0.49\textwidth]{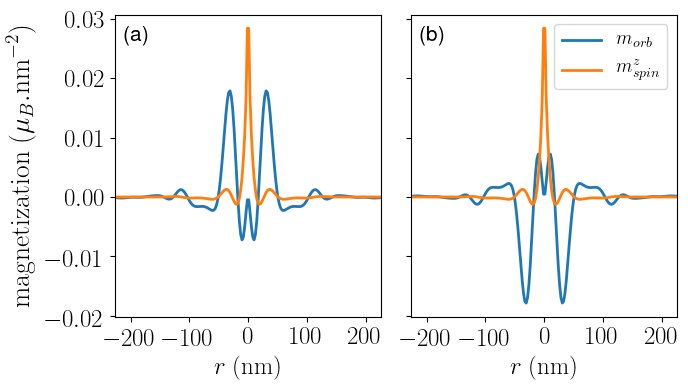}
  \caption{Orbital ($m_{orb}$) and spin ($m^{z}_{spin}$) contributions to the 2DEG magnetization density showing the effects of the current circulation inversion between $\tau = \pi/4 - 0.1$ (a) and $\tau = \pi/4 + 0.1$ (b) along the  $[1\overline{1}0]$ direction from  a magnetic defect pointing perpendicular to the 2DEG.}
\label{fig:2}
\end{figure}

The spatial structure of the orbital magnetization density associated with these circulating currents can be calculated in analogy with classical electrodynamics using\cite{Jackson} $\mathbf{m}_{orb}\left(\mathbf{r}\right)=(1/2)\mathbf{r}\times \mathbf{j}\left(\mathbf{r}\right)$; the current density is obtained from Eq.~(\ref{eq:current}). To compare both the orbital and spin contributions to the defect-induced magnetization we calculate the latter by

\begin{equation}
    \mathbf{m}_{spin}\left(\mathbf{r}\right)=-\frac{\mu_{B}}{\pi}\mathrm{Im}\int_{E_{so}}^{E_{f}}\bm{\sigma}G\left(\mathbf{r};E\right) dE,
\end{equation}
where  $\mu_{B}$ is the Bohr magneton and $\bm{\sigma}$ are the Pauli matrices. We note that orbital momentum quenching is reduced for systems with strong spin-orbit coupling, allowing a comparable  orbital magnetization to the spin magnetization. As shown in Fig.~\ref{fig:2}, although the spin-orbit fields alter both orbital and spin magnetization densities around a magnetic defect, the orbital magnetization has distinct spatial oscillations and is highly sensitive to the SOC ratio $\tau$.

The magnetic fringe field above the 2DEG is obtained from the orbital magnetization density of a quantum-well with width $d$ confined along the z-axis under hard-wall boundary conditions. By using the ground-state electron envelope function, $\phi\left(z\right) = (1/\sqrt{d})\cos\left(\pi z / d\right)$, the previously calculated two-dimensional current density acquires a z-dependence of $\cos^{2}\left(\pi z / d\right)$, and a volumetric orbital magnetization density can be defined inside the quantum well.

In the region outside the magnetized volume, the magnetic fringe field is evaluated by  $\mathbf{B}=-\mu_{0}\nabla\Phi_{orb}\left(\mathbf{r}\right)$, where $\mu_{0} = 4\pi \times 10^{-7}\ \mathrm{N/A}^{2}$ is the vacuum permeability and the scalar magnetic potential is\cite{Jackson}
\begin{equation}
\Phi_{orb}\left(\mathbf{r}\right)=\frac{1}{4\pi}\int d\mathbf{r}^{\prime}\frac{\nabla\cdot\mathbf{m}_{\text{orb}}\left(\mathbf{r}^{\prime}\right)}{\left|\mathbf{r}-\mathbf{r}^{\prime}\right|}.   
\end{equation}

\begin{figure}[ht]
  \includegraphics[width=0.49\textwidth]{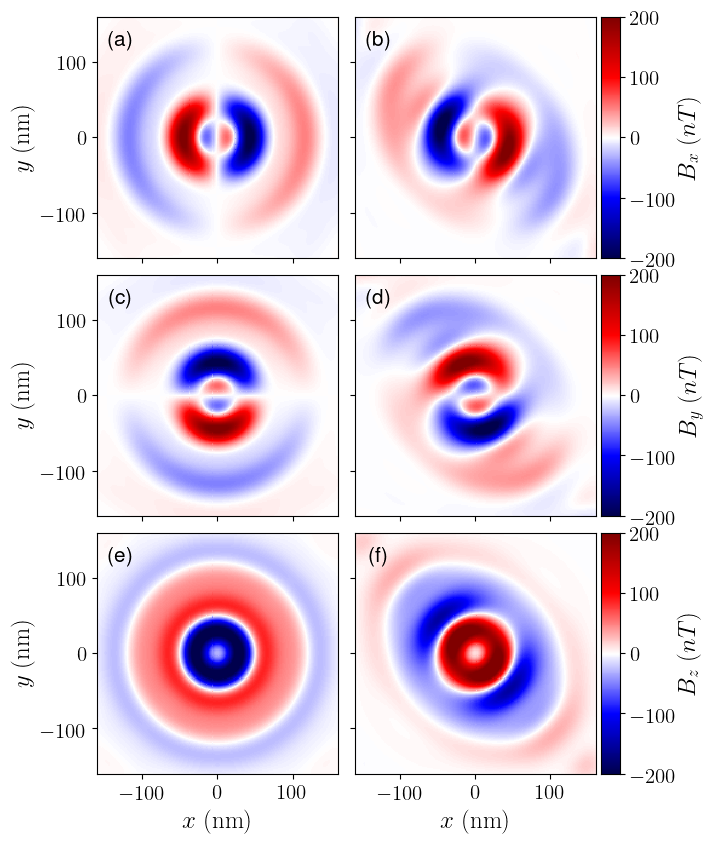}
  \caption{Orbital fringe field components (ab) $B_{x}$, (cd) $B_{y}$ and (ef) $B_{z}$  20 nm above the surface of an InAs quantum well with width 10~nm for a magnetic defect pointing perpendicular to the 2DEG. (ace) the SIA dominated regime,  $\tau = \pi/2$. (bdf)  for $\tau = \pi/8$. }
\label{fig:3}
\end{figure}
Figure~\ref{fig:3} shows the spatial distribution for each orbital fringe field component is calculated at 20 nm above a 10 nm InAs QW for this magnetic defect, and the largest values are tenths of $\mu T$, within the range of single NV diamond scanning magnetometry.  In the left column of Fig.~\ref{fig:3} the fringe field structure in the SIA dominated regime ($\tau = \pi/2$) reflects the radial symmetry of the circulating current in this regime, and the fields are generated by concentric  current loops with current flowing in opposite directions. The BIA term induces  angular anistropy in the field distribution [Fig.~\ref{fig:3}(bdf)] providing a direct signature of the SOC ratio.

\begin{figure}[ht]

  \includegraphics[width=0.48\textwidth]{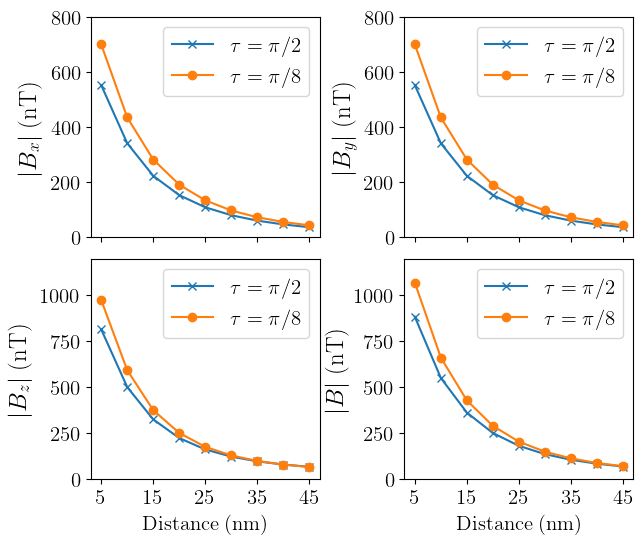}
  \caption{Maximum magnitude of the orbital fringe field components $B_{x}$, $B_{y}$ and $B_{z}$ and total magnetic field B calculated at different distances above a 10~nm InAs QW. Shown for $\beta=0$, $\tau=\pi/2$ (crosses and blue lines) $\beta\ne 0$, $\tau=\pi/8$ (circles and orange lines).}
\label{fig:4}
\end{figure}

The magnitude of the orbital fringe field depends on the distance measured from the QW, \textit{e.g.} decreasing fourfold at a distance of 20 nm from the InAs QW with respect to the field magnitude at the surface  (Fig.~\ref{fig:4}), thus the close proximity achievable by nanoscale scanning probes can obtain even stronger responses within its sensitivity range.  The higher magnitude shown for currents with both SIA and BIA terms (crosses and  blue lines in Fig.~\ref{fig:4}) occurs from higher currents due to the SOC-induced focusing effects. Furthermore, choices of different materials may lead to more detectable effects. Table~\ref{tab:table1} lists a variety of materials with either stronger magnetic fields or larger-scale spatial features, along with some more challenging materials that would have both (\textit{i.e.} BiSb). In Table~\ref{tab:table1} we keep the two-dimensional electron density fixed to be $4\times 10^{11}$~cm$^{-2}$, and note that the features can also become larger (or smaller) for different electron densities.

\begin{table}[h!]
\caption{\label{tab:table1}
Parameters and conversion factors for the spatial dimension ($d_\mathrm{fac}$) and magnetic field strength ($B_{\mathrm{fac}}$) for different materials.  We assume the QW width is $10\ \mathrm{nm}$.} 

\begin{ruledtabular}
\begin{tabular}{ccccc}
Material  & $m^{*} (m_0)$ & $\alpha$ (meV.nm) & $d_{\mathrm{fac}}$ & $B_{\mathrm{fac}}$ \\ [0.5ex] 
 
InAs
\cite{Grundler2000LargeLayers}  & 0.023  & 50    & 1.0 & 1.0 \\ 

GaAs/$\text{Al}\text{Ga}\text{As}$
\cite{Yu2016TuningWells}  & 0.067  & 4.8  & 3.58 & 0.096 \\ 

LAO/STO
\cite{Fete2014LargeInterface}
& 2.2  & 3.4  & 0.15  & 0.068\\ 

InGaAs/InAlAs
\cite{Nitta1997GateHeterostructure} & 0.05 & 40.0 & 0.57  & 0.8 \\ 

BiSb
\cite{Singh_BiSb} & 0.002 & 230 & 2.5 & 4.6 \\ 

LaOBiS\textsubscript{2} 
\cite{Liu2013TunableTransistors} & 0.07 & 478 & 0.034 & 9.56 \\ 

WSeTe  
\cite{Yao2017ManipulationDichalcogenides,Bahrami2021} & 0.81 & 92 & 0.015 & 1.84 \\ 

\end{tabular}
\end{ruledtabular}
\end{table}


The circulating dissipationless current associated with a single spin in a 2DEG formed in non-centrosymmetric materials with SOC creates spatial features highly sensitive to the underlying tunable spin-orbit field and the defect-spin direction, a reminiscent of the anisotropric dispersion relation and Friedel oscillations. Nanoscale scanning probe magnetometry, or potentially electron ptychography, is sufficiently sensitive to  measure the spatially-resolved orbital contribution to the  magnetic moment of a single spin. The spatial structure of the defect magnetic moment affects its coupling to nearby rapidly oscillating fields from e.g, nuclear-spins, with implications for  spin-dynamics and coherent control of single spin states \cite{VanBree2014Spin-orbit-inducedNanostructure}.

\begin{acknowledgments}
We acknowledge support from the ITN 4PHOTON
Marie Sklodowska Curie Grant Agreement No. 721394. 
\end{acknowledgments}

\bibliography{references}

\begin{thebibliography}{51}%
\makeatletter
\providecommand \@ifxundefined [1]{%
 \@ifx{#1\undefined}
}%
\providecommand \@ifnum [1]{%
 \ifnum #1\expandafter \@firstoftwo
 \else \expandafter \@secondoftwo
 \fi
}%
\providecommand \@ifx [1]{%
 \ifx #1\expandafter \@firstoftwo
 \else \expandafter \@secondoftwo
 \fi
}%
\providecommand \natexlab [1]{#1}%
\providecommand \enquote  [1]{``#1''}%
\providecommand \bibnamefont  [1]{#1}%
\providecommand \bibfnamefont [1]{#1}%
\providecommand \citenamefont [1]{#1}%
\providecommand \href@noop [0]{\@secondoftwo}%
\providecommand \href [0]{\begingroup \@sanitize@url \@href}%
\providecommand \@href[1]{\@@startlink{#1}\@@href}%
\providecommand \@@href[1]{\endgroup#1\@@endlink}%
\providecommand \@sanitize@url [0]{\catcode `\\12\catcode `\$12\catcode
  `\&12\catcode `\#12\catcode `\^12\catcode `\_12\catcode `\%12\relax}%
\providecommand \@@startlink[1]{}%
\providecommand \@@endlink[0]{}%
\providecommand \url  [0]{\begingroup\@sanitize@url \@url }%
\providecommand \@url [1]{\endgroup\@href {#1}{\urlprefix }}%
\providecommand \urlprefix  [0]{URL }%
\providecommand \Eprint [0]{\href }%
\providecommand \doibase [0]{https://doi.org/}%
\providecommand \selectlanguage [0]{\@gobble}%
\providecommand \bibinfo  [0]{\@secondoftwo}%
\providecommand \bibfield  [0]{\@secondoftwo}%
\providecommand \translation [1]{[#1]}%
\providecommand \BibitemOpen [0]{}%
\providecommand \bibitemStop [0]{}%
\providecommand \bibitemNoStop [0]{.\EOS\space}%
\providecommand \EOS [0]{\spacefactor3000\relax}%
\providecommand \BibitemShut  [1]{\csname bibitem#1\endcsname}%
\let\auto@bib@innerbib\@empty
\bibitem [{\citenamefont {Kane}(1998)}]{Kane1998AComputer}%
  \BibitemOpen
  \bibfield  {author} {\bibinfo {author} {\bibfnamefont {B.~E.}\ \bibnamefont
  {Kane}},\ }\href@noop {} {\bibfield  {journal} {\bibinfo  {journal} {Nature
  (London)}\ }\textbf {\bibinfo {volume} {393}},\ \bibinfo {pages} {133}
  (\bibinfo {year} {1998})}\BibitemShut {NoStop}%
\bibitem [{\citenamefont {Loss}\ and\ \citenamefont
  {DiVincenzo}(1998)}]{Loss1998}%
  \BibitemOpen
  \bibfield  {author} {\bibinfo {author} {\bibfnamefont {D.}~\bibnamefont
  {Loss}}\ and\ \bibinfo {author} {\bibfnamefont {D.~P.}\ \bibnamefont
  {DiVincenzo}},\ }\href {https://doi.org/10.1103/PhysRevA.57.120} {\bibfield
  {journal} {\bibinfo  {journal} {Phys. Rev. A}\ }\textbf {\bibinfo {volume}
  {57}},\ \bibinfo {pages} {120} (\bibinfo {year} {1998})}\BibitemShut
  {NoStop}%
\bibitem [{\citenamefont {Awschalom}\ \emph {et~al.}(2002)\citenamefont
  {Awschalom}, \citenamefont {Samarth},\ and\ \citenamefont
  {Loss}}]{Awschalom2002}%
  \BibitemOpen
  \bibinfo {editor} {\bibfnamefont {D.~D.}\ \bibnamefont {Awschalom}}, \bibinfo
  {editor} {\bibfnamefont {N.}~\bibnamefont {Samarth}},\ and\ \bibinfo {editor}
  {\bibfnamefont {D.}~\bibnamefont {Loss}},\ eds.,\ \href@noop {} {\emph
  {\bibinfo {title} {Semiconductor Spintronics and Quantum Computation}}}\
  (\bibinfo  {publisher} {Springer Verlag},\ \bibinfo {address} {Heidelberg},\
  \bibinfo {year} {2002})\BibitemShut {NoStop}%
\bibitem [{\citenamefont {Ichimura}(2001)}]{Ichimura2001AMode}%
  \BibitemOpen
  \bibfield  {author} {\bibinfo {author} {\bibfnamefont {K.}~\bibnamefont
  {Ichimura}},\ }\href {https://doi.org/10.1016/S0030-4018(01)01394-3}
  {\bibfield  {journal} {\bibinfo  {journal} {Opt. Commun.}\ }\textbf {\bibinfo
  {volume} {196}},\ \bibinfo {pages} {119} (\bibinfo {year}
  {2001})}\BibitemShut {NoStop}%
\bibitem [{\citenamefont {Wolfowicz}\ \emph {et~al.}(2021)\citenamefont
  {Wolfowicz}, \citenamefont {Heremans}, \citenamefont {Anderson},
  \citenamefont {Kanai}, \citenamefont {Seo}, \citenamefont {Gali},
  \citenamefont {Galli},\ and\ \citenamefont
  {Awschalom}}]{Wolfowicz2021QuantumDefects}%
  \BibitemOpen
  \bibfield  {author} {\bibinfo {author} {\bibfnamefont {G.}~\bibnamefont
  {Wolfowicz}}, \bibinfo {author} {\bibfnamefont {F.~J.}\ \bibnamefont
  {Heremans}}, \bibinfo {author} {\bibfnamefont {C.~P.}\ \bibnamefont
  {Anderson}}, \bibinfo {author} {\bibfnamefont {S.}~\bibnamefont {Kanai}},
  \bibinfo {author} {\bibfnamefont {H.}~\bibnamefont {Seo}}, \bibinfo {author}
  {\bibfnamefont {A.}~\bibnamefont {Gali}}, \bibinfo {author} {\bibfnamefont
  {G.}~\bibnamefont {Galli}},\ and\ \bibinfo {author} {\bibfnamefont {D.~D.}\
  \bibnamefont {Awschalom}},\ }\href {www.nature.com/natrevmats} {\bibfield
  {journal} {\bibinfo  {journal} {Nat. Rev. Mater.}\ }\textbf {\bibinfo
  {volume} {6}},\ \bibinfo {pages} {906} (\bibinfo {year} {2021})}\BibitemShut
  {NoStop}%
\bibitem [{\citenamefont {Matsui}\ \emph {et~al.}(2007)\citenamefont {Matsui},
  \citenamefont {Meyer}, \citenamefont {Sacharow}, \citenamefont {Wiebe},\ and\
  \citenamefont {Wiesendanger}}]{Matsui2007}%
  \BibitemOpen
  \bibfield  {author} {\bibinfo {author} {\bibfnamefont {T.}~\bibnamefont
  {Matsui}}, \bibinfo {author} {\bibfnamefont {C.}~\bibnamefont {Meyer}},
  \bibinfo {author} {\bibfnamefont {L.}~\bibnamefont {Sacharow}}, \bibinfo
  {author} {\bibfnamefont {J.}~\bibnamefont {Wiebe}},\ and\ \bibinfo {author}
  {\bibfnamefont {R.}~\bibnamefont {Wiesendanger}},\ }\href
  {https://doi.org/10.1103/PhysRevB.75.165405} {\bibfield  {journal} {\bibinfo
  {journal} {Phys. Rev. B}\ }\textbf {\bibinfo {volume} {75}},\ \bibinfo
  {pages} {165405} (\bibinfo {year} {2007})}\BibitemShut {NoStop}%
\bibitem [{\citenamefont {Lee}\ and\ \citenamefont {Gupta}(2010)}]{Lee2010}%
  \BibitemOpen
  \bibfield  {author} {\bibinfo {author} {\bibfnamefont {D.~H.}\ \bibnamefont
  {Lee}}\ and\ \bibinfo {author} {\bibfnamefont {J.~A.}\ \bibnamefont
  {Gupta}},\ }\href@noop {} {\bibfield  {journal} {\bibinfo  {journal}
  {Science}\ }\textbf {\bibinfo {volume} {330}},\ \bibinfo {pages} {1807}
  (\bibinfo {year} {2010})}\BibitemShut {NoStop}%
\bibitem [{\citenamefont {Myers}\ \emph {et~al.}(2008)\citenamefont {Myers},
  \citenamefont {Mikkelsen}, \citenamefont {Tang}, \citenamefont {Gossard},
  \citenamefont {Flatt{\'{e}}},\ and\ \citenamefont
  {Awschalom}}]{Myers2008Zero-fieldSemiconductors}%
  \BibitemOpen
  \bibfield  {author} {\bibinfo {author} {\bibfnamefont {R.~C.}\ \bibnamefont
  {Myers}}, \bibinfo {author} {\bibfnamefont {M.~H.}\ \bibnamefont
  {Mikkelsen}}, \bibinfo {author} {\bibfnamefont {J.-M.}\ \bibnamefont {Tang}},
  \bibinfo {author} {\bibfnamefont {A.~C.}\ \bibnamefont {Gossard}}, \bibinfo
  {author} {\bibfnamefont {M.~E.}\ \bibnamefont {Flatt{\'{e}}}},\ and\ \bibinfo
  {author} {\bibfnamefont {D.~D.}\ \bibnamefont {Awschalom}},\ }\href@noop {}
  {\bibfield  {journal} {\bibinfo  {journal} {Nat. Mater.}\ }\textbf {\bibinfo
  {volume} {7}},\ \bibinfo {pages} {203} (\bibinfo {year} {2008})}\BibitemShut
  {NoStop}%
\bibitem [{\citenamefont {Gaebel}\ \emph {et~al.}(2016)\citenamefont {Gaebel},
  \citenamefont {Domhan}, \citenamefont {Popa}, \citenamefont {Wittmann},
  \citenamefont {Neumann}, \citenamefont {Jelezko}, \citenamefont {Rabeau},
  \citenamefont {Stavrias}, \citenamefont {Greentree}, \citenamefont {Prawer},
  \citenamefont {Meijer}, \citenamefont {Twamley}, \citenamefont {Hemmer},\
  and\ \citenamefont {Wrachtrup}}]{Gaebel2006}%
  \BibitemOpen
  \bibfield  {author} {\bibinfo {author} {\bibfnamefont {T.}~\bibnamefont
  {Gaebel}}, \bibinfo {author} {\bibfnamefont {M.}~\bibnamefont {Domhan}},
  \bibinfo {author} {\bibfnamefont {I.}~\bibnamefont {Popa}}, \bibinfo {author}
  {\bibfnamefont {C.}~\bibnamefont {Wittmann}}, \bibinfo {author}
  {\bibfnamefont {P.}~\bibnamefont {Neumann}}, \bibinfo {author} {\bibfnamefont
  {F.}~\bibnamefont {Jelezko}}, \bibinfo {author} {\bibfnamefont {J.~R.}\
  \bibnamefont {Rabeau}}, \bibinfo {author} {\bibfnamefont {N.}~\bibnamefont
  {Stavrias}}, \bibinfo {author} {\bibfnamefont {A.~D.}\ \bibnamefont
  {Greentree}}, \bibinfo {author} {\bibfnamefont {S.}~\bibnamefont {Prawer}},
  \bibinfo {author} {\bibfnamefont {J.}~\bibnamefont {Meijer}}, \bibinfo
  {author} {\bibfnamefont {J.}~\bibnamefont {Twamley}}, \bibinfo {author}
  {\bibfnamefont {P.~R.}\ \bibnamefont {Hemmer}},\ and\ \bibinfo {author}
  {\bibfnamefont {J.}~\bibnamefont {Wrachtrup}},\ }\href
  {https://www.nature.com/articles/nphys318} {\bibfield  {journal} {\bibinfo
  {journal} {Nature Phys}\ }\textbf {\bibinfo {volume} {2}},\ \bibinfo {pages}
  {408} (\bibinfo {year} {2016})}\BibitemShut {NoStop}%
\bibitem [{\citenamefont {Fuchs}\ \emph {et~al.}(2011)\citenamefont {Fuchs},
  \citenamefont {Burkard}, \citenamefont {Klimov},\ and\ \citenamefont
  {Awschalom}}]{Fuchs2011}%
  \BibitemOpen
  \bibfield  {author} {\bibinfo {author} {\bibfnamefont {G.~D.}\ \bibnamefont
  {Fuchs}}, \bibinfo {author} {\bibfnamefont {G.}~\bibnamefont {Burkard}},
  \bibinfo {author} {\bibfnamefont {P.~V.}\ \bibnamefont {Klimov}},\ and\
  \bibinfo {author} {\bibfnamefont {D.~D.}\ \bibnamefont {Awschalom}},\ }\href
  {https://www.nature.com/articles/nphys2026} {\bibfield  {journal} {\bibinfo
  {journal} {Nature Phys}\ }\textbf {\bibinfo {volume} {7}},\ \bibinfo {pages}
  {789} (\bibinfo {year} {2011})}\BibitemShut {NoStop}%
\bibitem [{\citenamefont {Togan}\ \emph {et~al.}(2010)\citenamefont {Togan},
  \citenamefont {Chu}, \citenamefont {Trifonov}, \citenamefont {Jiang},
  \citenamefont {Maze}, \citenamefont {Childress}, \citenamefont {Dutt},
  \citenamefont {Sørensen}, \citenamefont {Hemmer}, \citenamefont {Zibrov},\
  and\ \citenamefont {Lukin}}]{Togan2010}%
  \BibitemOpen
  \bibfield  {author} {\bibinfo {author} {\bibfnamefont {E.}~\bibnamefont
  {Togan}}, \bibinfo {author} {\bibfnamefont {Y.}~\bibnamefont {Chu}}, \bibinfo
  {author} {\bibfnamefont {A.~S.}\ \bibnamefont {Trifonov}}, \bibinfo {author}
  {\bibfnamefont {L.}~\bibnamefont {Jiang}}, \bibinfo {author} {\bibfnamefont
  {J.}~\bibnamefont {Maze}}, \bibinfo {author} {\bibfnamefont {L.}~\bibnamefont
  {Childress}}, \bibinfo {author} {\bibfnamefont {M.~V.~G.}\ \bibnamefont
  {Dutt}}, \bibinfo {author} {\bibfnamefont {A.~S.}\ \bibnamefont {Sørensen}},
  \bibinfo {author} {\bibfnamefont {P.~R.}\ \bibnamefont {Hemmer}}, \bibinfo
  {author} {\bibfnamefont {A.~S.}\ \bibnamefont {Zibrov}},\ and\ \bibinfo
  {author} {\bibfnamefont {M.~D.}\ \bibnamefont {Lukin}},\ }\href
  {https://www.nature.com/articles/nature09256} {\bibfield  {journal} {\bibinfo
   {journal} {Nature}\ }\textbf {\bibinfo {volume} {466}},\ \bibinfo {pages}
  {730} (\bibinfo {year} {2010})}\BibitemShut {NoStop}%
\bibitem [{\citenamefont {Kato}\ \emph {et~al.}(2003)\citenamefont {Kato},
  \citenamefont {Myers}, \citenamefont {Driscoll}, \citenamefont {Gossard},
  \citenamefont {Levy},\ and\ \citenamefont {Awschalom}}]{Kato2003}%
  \BibitemOpen
  \bibfield  {author} {\bibinfo {author} {\bibfnamefont {Y.}~\bibnamefont
  {Kato}}, \bibinfo {author} {\bibfnamefont {R.~C.}\ \bibnamefont {Myers}},
  \bibinfo {author} {\bibfnamefont {D.~C.}\ \bibnamefont {Driscoll}}, \bibinfo
  {author} {\bibfnamefont {A.~C.}\ \bibnamefont {Gossard}}, \bibinfo {author}
  {\bibfnamefont {J.}~\bibnamefont {Levy}},\ and\ \bibinfo {author}
  {\bibfnamefont {D.~D.}\ \bibnamefont {Awschalom}},\ }\href
  {https://doi.org/10.1126/SCIENCE.1080880} {\bibfield  {journal} {\bibinfo
  {journal} {Science}\ }\textbf {\bibinfo {volume} {299}},\ \bibinfo {pages}
  {1201} (\bibinfo {year} {2003})}\BibitemShut {NoStop}%
\bibitem [{\citenamefont {Pingenot}\ \emph {et~al.}(2008)\citenamefont
  {Pingenot}, \citenamefont {Pryor},\ and\ \citenamefont
  {Flatté}}]{Pingenot2008}%
  \BibitemOpen
  \bibfield  {author} {\bibinfo {author} {\bibfnamefont {J.}~\bibnamefont
  {Pingenot}}, \bibinfo {author} {\bibfnamefont {C.~E.}\ \bibnamefont
  {Pryor}},\ and\ \bibinfo {author} {\bibfnamefont {M.~E.}\ \bibnamefont
  {Flatté}},\ }\href@noop {} {\bibfield  {journal} {\bibinfo  {journal}
  {Applied Physics Letters}\ }\textbf {\bibinfo {volume} {92}},\ \bibinfo
  {pages} {222502} (\bibinfo {year} {2008})}\BibitemShut {NoStop}%
\bibitem [{\citenamefont {Pingenot}\ \emph {et~al.}(2011)\citenamefont
  {Pingenot}, \citenamefont {Pryor},\ and\ \citenamefont
  {Flatt\'e}}]{Pingenot2011}%
  \BibitemOpen
  \bibfield  {author} {\bibinfo {author} {\bibfnamefont {J.}~\bibnamefont
  {Pingenot}}, \bibinfo {author} {\bibfnamefont {C.~E.}\ \bibnamefont
  {Pryor}},\ and\ \bibinfo {author} {\bibfnamefont {M.~E.}\ \bibnamefont
  {Flatt\'e}},\ }\href {https://doi.org/10.1103/PhysRevB.84.195403} {\bibfield
  {journal} {\bibinfo  {journal} {Phys. Rev. B}\ }\textbf {\bibinfo {volume}
  {84}},\ \bibinfo {pages} {195403} (\bibinfo {year} {2011})}\BibitemShut
  {NoStop}%
\bibitem [{\citenamefont {Schroer}\ \emph {et~al.}(2011)\citenamefont
  {Schroer}, \citenamefont {Petersson}, \citenamefont {Jung},\ and\
  \citenamefont {Petta}}]{Schroer2011}%
  \BibitemOpen
  \bibfield  {author} {\bibinfo {author} {\bibfnamefont {M.~D.}\ \bibnamefont
  {Schroer}}, \bibinfo {author} {\bibfnamefont {K.~D.}\ \bibnamefont
  {Petersson}}, \bibinfo {author} {\bibfnamefont {M.}~\bibnamefont {Jung}},\
  and\ \bibinfo {author} {\bibfnamefont {J.~R.}\ \bibnamefont {Petta}},\ }\href
  {https://doi.org/10.1103/PhysRevLett.107.176811} {\bibfield  {journal}
  {\bibinfo  {journal} {Phys. Rev. Lett.}\ }\textbf {\bibinfo {volume} {107}},\
  \bibinfo {pages} {176811} (\bibinfo {year} {2011})}\BibitemShut {NoStop}%
\bibitem [{\citenamefont {Takahashi}\ \emph {et~al.}(2013)\citenamefont
  {Takahashi}, \citenamefont {Deacon}, \citenamefont {Oiwa}, \citenamefont
  {Shibata}, \citenamefont {Hirakawa},\ and\ \citenamefont
  {Tarucha}}]{Takahashi2013}%
  \BibitemOpen
  \bibfield  {author} {\bibinfo {author} {\bibfnamefont {S.}~\bibnamefont
  {Takahashi}}, \bibinfo {author} {\bibfnamefont {R.~S.}\ \bibnamefont
  {Deacon}}, \bibinfo {author} {\bibfnamefont {A.}~\bibnamefont {Oiwa}},
  \bibinfo {author} {\bibfnamefont {K.}~\bibnamefont {Shibata}}, \bibinfo
  {author} {\bibfnamefont {K.}~\bibnamefont {Hirakawa}},\ and\ \bibinfo
  {author} {\bibfnamefont {S.}~\bibnamefont {Tarucha}},\ }\href
  {https://doi.org/10.1103/PhysRevB.87.161302} {\bibfield  {journal} {\bibinfo
  {journal} {Phys. Rev. B}\ }\textbf {\bibinfo {volume} {87}},\ \bibinfo
  {pages} {161302(R)} (\bibinfo {year} {2013})}\BibitemShut {NoStop}%
\bibitem [{\citenamefont {Ares}\ \emph {et~al.}(2013)\citenamefont {Ares},
  \citenamefont {Katsaros}, \citenamefont {Golovach}, \citenamefont {Zhang},
  \citenamefont {Prager}, \citenamefont {Glazman}, \citenamefont {Schmidt},\
  and\ \citenamefont {De~Franceschi}}]{Ares2013}%
  \BibitemOpen
  \bibfield  {author} {\bibinfo {author} {\bibfnamefont {N.}~\bibnamefont
  {Ares}}, \bibinfo {author} {\bibfnamefont {G.}~\bibnamefont {Katsaros}},
  \bibinfo {author} {\bibfnamefont {V.~N.}\ \bibnamefont {Golovach}}, \bibinfo
  {author} {\bibfnamefont {J.~J.}\ \bibnamefont {Zhang}}, \bibinfo {author}
  {\bibfnamefont {A.}~\bibnamefont {Prager}}, \bibinfo {author} {\bibfnamefont
  {L.~I.}\ \bibnamefont {Glazman}}, \bibinfo {author} {\bibfnamefont {O.~G.}\
  \bibnamefont {Schmidt}},\ and\ \bibinfo {author} {\bibfnamefont
  {S.}~\bibnamefont {De~Franceschi}},\ }\href@noop {} {\bibfield  {journal}
  {\bibinfo  {journal} {Applied Physics Letters}\ }\textbf {\bibinfo {volume}
  {103}},\ \bibinfo {pages} {263113} (\bibinfo {year} {2013})}\BibitemShut
  {NoStop}%
\bibitem [{\citenamefont {Crippa}\ \emph {et~al.}(2018)\citenamefont {Crippa},
  \citenamefont {Maurand}, \citenamefont {Bourdet}, \citenamefont
  {Kotekar-Patil}, \citenamefont {Amisse}, \citenamefont {Jehl}, \citenamefont
  {Sanquer}, \citenamefont {Lavi\'eville}, \citenamefont {Bohuslavskyi},
  \citenamefont {Hutin}, \citenamefont {Barraud}, \citenamefont {Vinet},
  \citenamefont {Niquet},\ and\ \citenamefont {De~Franceschi}}]{Crippa2018}%
  \BibitemOpen
  \bibfield  {author} {\bibinfo {author} {\bibfnamefont {A.}~\bibnamefont
  {Crippa}}, \bibinfo {author} {\bibfnamefont {R.}~\bibnamefont {Maurand}},
  \bibinfo {author} {\bibfnamefont {L.}~\bibnamefont {Bourdet}}, \bibinfo
  {author} {\bibfnamefont {D.}~\bibnamefont {Kotekar-Patil}}, \bibinfo {author}
  {\bibfnamefont {A.}~\bibnamefont {Amisse}}, \bibinfo {author} {\bibfnamefont
  {X.}~\bibnamefont {Jehl}}, \bibinfo {author} {\bibfnamefont {M.}~\bibnamefont
  {Sanquer}}, \bibinfo {author} {\bibfnamefont {R.}~\bibnamefont
  {Lavi\'eville}}, \bibinfo {author} {\bibfnamefont {H.}~\bibnamefont
  {Bohuslavskyi}}, \bibinfo {author} {\bibfnamefont {L.}~\bibnamefont {Hutin}},
  \bibinfo {author} {\bibfnamefont {S.}~\bibnamefont {Barraud}}, \bibinfo
  {author} {\bibfnamefont {M.}~\bibnamefont {Vinet}}, \bibinfo {author}
  {\bibfnamefont {Y.-M.}\ \bibnamefont {Niquet}},\ and\ \bibinfo {author}
  {\bibfnamefont {S.}~\bibnamefont {De~Franceschi}},\ }\href
  {https://doi.org/10.1103/PhysRevLett.120.137702} {\bibfield  {journal}
  {\bibinfo  {journal} {Phys. Rev. Lett.}\ }\textbf {\bibinfo {volume} {120}},\
  \bibinfo {pages} {137702} (\bibinfo {year} {2018})}\BibitemShut {NoStop}%
\bibitem [{\citenamefont {Ferr\'on}\ \emph {et~al.}(2019)\citenamefont
  {Ferr\'on}, \citenamefont {Rodr\'{\i}guez}, \citenamefont {G\'omez},
  \citenamefont {Lado},\ and\ \citenamefont
  {Fern\'andez-Rossier}}]{Ferron2019}%
  \BibitemOpen
  \bibfield  {author} {\bibinfo {author} {\bibfnamefont {A.}~\bibnamefont
  {Ferr\'on}}, \bibinfo {author} {\bibfnamefont {S.~A.}\ \bibnamefont
  {Rodr\'{\i}guez}}, \bibinfo {author} {\bibfnamefont {S.~S.}\ \bibnamefont
  {G\'omez}}, \bibinfo {author} {\bibfnamefont {J.~L.}\ \bibnamefont {Lado}},\
  and\ \bibinfo {author} {\bibfnamefont {J.}~\bibnamefont
  {Fern\'andez-Rossier}},\ }\href
  {https://doi.org/10.1103/PhysRevResearch.1.033185} {\bibfield  {journal}
  {\bibinfo  {journal} {Phys. Rev. Research}\ }\textbf {\bibinfo {volume}
  {1}},\ \bibinfo {pages} {033185} (\bibinfo {year} {2019})}\BibitemShut
  {NoStop}%
\bibitem [{\citenamefont {Doherty}\ \emph {et~al.}(2013)\citenamefont
  {Doherty}, \citenamefont {Manson}, \citenamefont {Delaney}, \citenamefont
  {Jelezko}, \citenamefont {Wrachtrup},\ and\ \citenamefont
  {Hollenberg}}]{Doherty2013TheDiamond}%
  \BibitemOpen
  \bibfield  {author} {\bibinfo {author} {\bibfnamefont {M.~W.}\ \bibnamefont
  {Doherty}}, \bibinfo {author} {\bibfnamefont {N.~B.}\ \bibnamefont {Manson}},
  \bibinfo {author} {\bibfnamefont {P.}~\bibnamefont {Delaney}}, \bibinfo
  {author} {\bibfnamefont {F.}~\bibnamefont {Jelezko}}, \bibinfo {author}
  {\bibfnamefont {J.}~\bibnamefont {Wrachtrup}},\ and\ \bibinfo {author}
  {\bibfnamefont {L.~C.}\ \bibnamefont {Hollenberg}},\ }\href
  {https://doi.org/10.1016/j.physrep.2013.02.001} {\bibfield  {journal}
  {\bibinfo  {journal} {Physics Reports}\ }\textbf {\bibinfo {volume} {528}},\
  \bibinfo {pages} {1} (\bibinfo {year} {2013})}\BibitemShut {NoStop}%
\bibitem [{\citenamefont {Grinolds}\ \emph {et~al.}(2013)\citenamefont
  {Grinolds}, \citenamefont {Hong}, \citenamefont {Maletinsky}, \citenamefont
  {Luan}, \citenamefont {Lukin}, \citenamefont {Walsworth},\ and\ \citenamefont
  {Yacoby}}]{Grinolds2013NanoscaleConditions}%
  \BibitemOpen
  \bibfield  {author} {\bibinfo {author} {\bibfnamefont {M.~S.}\ \bibnamefont
  {Grinolds}}, \bibinfo {author} {\bibfnamefont {S.}~\bibnamefont {Hong}},
  \bibinfo {author} {\bibfnamefont {P.}~\bibnamefont {Maletinsky}}, \bibinfo
  {author} {\bibfnamefont {L.}~\bibnamefont {Luan}}, \bibinfo {author}
  {\bibfnamefont {M.~D.}\ \bibnamefont {Lukin}}, \bibinfo {author}
  {\bibfnamefont {R.~L.}\ \bibnamefont {Walsworth}},\ and\ \bibinfo {author}
  {\bibfnamefont {A.}~\bibnamefont {Yacoby}},\ }\href
  {www.nature.com/naturephysics} {\bibfield  {journal} {\bibinfo  {journal}
  {Nat. Phys. Lett.}\ }\textbf {\bibinfo {volume} {9}},\ \bibinfo {pages} {215}
  (\bibinfo {year} {2013})}\BibitemShut {NoStop}%
\bibitem [{\citenamefont {Casola}\ \emph {et~al.}(2018)\citenamefont {Casola},
  \citenamefont {Van Der~Sar},\ and\ \citenamefont
  {Yacoby}}]{Casola2018ProbingDiamond}%
  \BibitemOpen
  \bibfield  {author} {\bibinfo {author} {\bibfnamefont {F.}~\bibnamefont
  {Casola}}, \bibinfo {author} {\bibfnamefont {T.}~\bibnamefont {Van
  Der~Sar}},\ and\ \bibinfo {author} {\bibfnamefont {A.}~\bibnamefont
  {Yacoby}},\ }\href {https://doi.org/10.1038/natrevmats.2017.88} {\bibfield
  {journal} {\bibinfo  {journal} {Nat. Rev. Mater.}\ }\textbf {\bibinfo
  {volume} {3}},\ \bibinfo {pages} {17088} (\bibinfo {year}
  {2018})}\BibitemShut {NoStop}%
\bibitem [{\citenamefont {Degen}\ \emph {et~al.}(2017)\citenamefont {Degen},
  \citenamefont {Reinhard},\ and\ \citenamefont {Cappellaro}}]{Degen2017}%
  \BibitemOpen
  \bibfield  {author} {\bibinfo {author} {\bibfnamefont {C.~L.}\ \bibnamefont
  {Degen}}, \bibinfo {author} {\bibfnamefont {F.}~\bibnamefont {Reinhard}},\
  and\ \bibinfo {author} {\bibfnamefont {P.}~\bibnamefont {Cappellaro}},\
  }\href {https://doi.org/10.1103/RevModPhys.89.035002} {\bibfield  {journal}
  {\bibinfo  {journal} {Rev. Mod. Phys.}\ }\textbf {\bibinfo {volume} {89}},\
  \bibinfo {pages} {035002} (\bibinfo {year} {2017})}\BibitemShut {NoStop}%
\bibitem [{\citenamefont {Noguchi}\ \emph {et~al.}(1991)\citenamefont
  {Noguchi}, \citenamefont {Hirakawa},\ and\ \citenamefont
  {Ikoma}}]{Noguchi1991InMinato}%
  \BibitemOpen
  \bibfield  {author} {\bibinfo {author} {\bibfnamefont {M.}~\bibnamefont
  {Noguchi}}, \bibinfo {author} {\bibfnamefont {K.}~\bibnamefont {Hirakawa}},\
  and\ \bibinfo {author} {\bibfnamefont {T.}~\bibnamefont {Ikoma}},\
  }\href@noop {} {\bibfield  {journal} {\bibinfo  {journal} {Phys. Rev. Lett.}\
  }\textbf {\bibinfo {volume} {66}},\ \bibinfo {pages} {2243} (\bibinfo {year}
  {1991})}\BibitemShut {NoStop}%
\bibitem [{\citenamefont {Bell}\ \emph {et~al.}(1997)\citenamefont {Bell},
  \citenamefont {Jones},\ and\ \citenamefont
  {Mcconville}}]{Bell1997AccumulationSurfaces}%
  \BibitemOpen
  \bibfield  {author} {\bibinfo {author} {\bibfnamefont {G.~R.}\ \bibnamefont
  {Bell}}, \bibinfo {author} {\bibfnamefont {T.~S.}\ \bibnamefont {Jones}},\
  and\ \bibinfo {author} {\bibfnamefont {C.~F.}\ \bibnamefont {Mcconville}},\
  }\href {https://doi.org/10.1063/1.120482} {\bibfield  {journal} {\bibinfo
  {journal} {Appl. Phys. Lett}\ }\textbf {\bibinfo {volume} {71}},\ \bibinfo
  {pages} {3688} (\bibinfo {year} {1997})}\BibitemShut {NoStop}%
\bibitem [{\citenamefont {Canali}\ \emph {et~al.}(1998)\citenamefont {Canali},
  \citenamefont {Wild{\"{o}}er}, \citenamefont {Kerkhof},\ and\ \citenamefont
  {Kouwenhoven}}]{Canali1998Low-temperatureSurfaces}%
  \BibitemOpen
  \bibfield  {author} {\bibinfo {author} {\bibfnamefont {L.}~\bibnamefont
  {Canali}}, \bibinfo {author} {\bibfnamefont {J.~W.~G.}\ \bibnamefont
  {Wild{\"{o}}er}}, \bibinfo {author} {\bibfnamefont {O.}~\bibnamefont
  {Kerkhof}},\ and\ \bibinfo {author} {\bibfnamefont {L.~P.}\ \bibnamefont
  {Kouwenhoven}},\ }\href@noop {} {\bibfield  {journal} {\bibinfo  {journal}
  {Appl. Phys. A}\ }\textbf {\bibinfo {volume} {66}},\ \bibinfo {pages} {113}
  (\bibinfo {year} {1998})}\BibitemShut {NoStop}%
\bibitem [{\citenamefont {F{\^{e}}te}\ \emph {et~al.}(2014)\citenamefont
  {F{\^{e}}te}, \citenamefont {Gariglio}, \citenamefont {Berthod},
  \citenamefont {Li}, \citenamefont {Stornaiuolo}, \citenamefont {Gabay},\ and\
  \citenamefont {Triscone}}]{Fete2014LargeInterface}%
  \BibitemOpen
  \bibfield  {author} {\bibinfo {author} {\bibfnamefont {A.}~\bibnamefont
  {F{\^{e}}te}}, \bibinfo {author} {\bibfnamefont {S.}~\bibnamefont
  {Gariglio}}, \bibinfo {author} {\bibfnamefont {C.}~\bibnamefont {Berthod}},
  \bibinfo {author} {\bibfnamefont {D.}~\bibnamefont {Li}}, \bibinfo {author}
  {\bibfnamefont {D.}~\bibnamefont {Stornaiuolo}}, \bibinfo {author}
  {\bibfnamefont {M.}~\bibnamefont {Gabay}},\ and\ \bibinfo {author}
  {\bibfnamefont {J.~M.}\ \bibnamefont {Triscone}},\ }\href
  {https://doi.org/10.1088/1367-2630/16/11/112002} {\bibfield  {journal}
  {\bibinfo  {journal} {New J. Phys.}\ }\textbf {\bibinfo {volume} {16}},\
  \bibinfo {pages} {112002} (\bibinfo {year} {2014})}\BibitemShut {NoStop}%
\bibitem [{\citenamefont {Singh}\ and\ \citenamefont
  {Romero}(2017)}]{Singh_BiSb}%
  \BibitemOpen
  \bibfield  {author} {\bibinfo {author} {\bibfnamefont {S.}~\bibnamefont
  {Singh}}\ and\ \bibinfo {author} {\bibfnamefont {A.~H.}\ \bibnamefont
  {Romero}},\ }\href {https://doi.org/10.1103/PhysRevB.95.165444} {\bibfield
  {journal} {\bibinfo  {journal} {Phys. Rev. B.}\ }\textbf {\bibinfo {volume}
  {95}},\ \bibinfo {pages} {165444} (\bibinfo {year} {2017})}\BibitemShut
  {NoStop}%
\bibitem [{\citenamefont {Yao}\ \emph {et~al.}(2017)\citenamefont {Yao},
  \citenamefont {Cai}, \citenamefont {Tong}, \citenamefont {Gong},
  \citenamefont {Wang}, \citenamefont {Wan}, \citenamefont {Duan},\ and\
  \citenamefont {Chu}}]{Yao2017ManipulationDichalcogenides}%
  \BibitemOpen
  \bibfield  {author} {\bibinfo {author} {\bibfnamefont {Q.-F.}\ \bibnamefont
  {Yao}}, \bibinfo {author} {\bibfnamefont {J.}~\bibnamefont {Cai}}, \bibinfo
  {author} {\bibfnamefont {W.-Y.}\ \bibnamefont {Tong}}, \bibinfo {author}
  {\bibfnamefont {S.-J.}\ \bibnamefont {Gong}}, \bibinfo {author}
  {\bibfnamefont {J.-Q.}\ \bibnamefont {Wang}}, \bibinfo {author}
  {\bibfnamefont {X.}~\bibnamefont {Wan}}, \bibinfo {author} {\bibfnamefont
  {C.-G.}\ \bibnamefont {Duan}},\ and\ \bibinfo {author} {\bibfnamefont
  {J.~H.}\ \bibnamefont {Chu}},\ }\href
  {https://doi.org/10.1103/PhysRevB.95.165401} {\bibfield  {journal} {\bibinfo
  {journal} {Phys. Rev. B}\ }\textbf {\bibinfo {volume} {95}},\ \bibinfo
  {pages} {165401} (\bibinfo {year} {2017})}\BibitemShut {NoStop}%
\bibitem [{\citenamefont {Liu}\ \emph {et~al.}(2013)\citenamefont {Liu},
  \citenamefont {Guo},\ and\ \citenamefont
  {Freeman}}]{Liu2013TunableTransistors}%
  \BibitemOpen
  \bibfield  {author} {\bibinfo {author} {\bibfnamefont {Q.}~\bibnamefont
  {Liu}}, \bibinfo {author} {\bibfnamefont {Y.}~\bibnamefont {Guo}},\ and\
  \bibinfo {author} {\bibfnamefont {A.~J.}\ \bibnamefont {Freeman}},\ }\href
  {https://doi.org/10.1021/nl4027346} {\bibfield  {journal} {\bibinfo
  {journal} {Nano Lett.}\ }\textbf {\bibinfo {volume} {13}},\ \bibinfo {pages}
  {5264} (\bibinfo {year} {2013})}\BibitemShut {NoStop}%
\bibitem [{\citenamefont {Friedel}(1958)}]{Friedel1958MetallicAlloys}%
  \BibitemOpen
  \bibfield  {author} {\bibinfo {author} {\bibfnamefont {J.}~\bibnamefont
  {Friedel}},\ }\href@noop {} {\bibfield  {journal} {\bibinfo  {journal} {Nuovo
  Cimento}\ }\textbf {\bibinfo {volume} {2}},\ \bibinfo {pages} {287} (\bibinfo
  {year} {1958})}\BibitemShut {NoStop}%
\bibitem [{\citenamefont {Lounis}\ \emph {et~al.}(2012)\citenamefont {Lounis},
  \citenamefont {Bringer},\ and\ \citenamefont
  {Bl{\"{u}}gel}}]{Lounis2012MagneticWaves}%
  \BibitemOpen
  \bibfield  {author} {\bibinfo {author} {\bibfnamefont {S.}~\bibnamefont
  {Lounis}}, \bibinfo {author} {\bibfnamefont {A.}~\bibnamefont {Bringer}},\
  and\ \bibinfo {author} {\bibfnamefont {S.}~\bibnamefont {Bl{\"{u}}gel}},\
  }\href {https://doi.org/10.1103/PhysRevLett.108.207202} {\bibfield  {journal}
  {\bibinfo  {journal} {Phys. Rev. Lett.}\ }\textbf {\bibinfo {volume} {108}},\
  \bibinfo {pages} {207202} (\bibinfo {year} {2012})}\BibitemShut {NoStop}%
\bibitem [{\citenamefont {Freimuth}\ \emph {et~al.}(2017)\citenamefont
  {Freimuth}, \citenamefont {Bl\"ugel},\ and\ \citenamefont
  {Mokrousov}}]{Freimuth2017}%
  \BibitemOpen
  \bibfield  {author} {\bibinfo {author} {\bibfnamefont {F.}~\bibnamefont
  {Freimuth}}, \bibinfo {author} {\bibfnamefont {S.}~\bibnamefont {Bl\"ugel}},\
  and\ \bibinfo {author} {\bibfnamefont {Y.}~\bibnamefont {Mokrousov}},\
  }\href@noop {} {\bibfield  {journal} {\bibinfo  {journal} {Phys. Rev. B}\
  }\textbf {\bibinfo {volume} {96}},\ \bibinfo {pages} {054403} (\bibinfo
  {year} {2017})}\BibitemShut {NoStop}%
\bibitem [{\citenamefont {Ruderman}\ and\ \citenamefont
  {Kittel}(1954)}]{Ruderman1954IndirectElectrons}%
  \BibitemOpen
  \bibfield  {author} {\bibinfo {author} {\bibfnamefont {M.~A.}\ \bibnamefont
  {Ruderman}}\ and\ \bibinfo {author} {\bibfnamefont {C.}~\bibnamefont
  {Kittel}},\ }\href@noop {} {\bibfield  {journal} {\bibinfo  {journal} {Phys.
  Rev.}\ }\textbf {\bibinfo {volume} {96}},\ \bibinfo {pages} {99} (\bibinfo
  {year} {1954})}\BibitemShut {NoStop}%
\bibitem [{\citenamefont {Kasuy}(1956)}]{Kasuy1956AModel}%
  \BibitemOpen
  \bibfield  {author} {\bibinfo {author} {\bibfnamefont {T.}~\bibnamefont
  {Kasuy}},\ }\href {https://academic.oup.com/ptp/article/16/1/45/1861363}
  {\bibfield  {journal} {\bibinfo  {journal} {Prog. Theor. Phys.}\ }\textbf
  {\bibinfo {volume} {16}},\ \bibinfo {pages} {45} (\bibinfo {year}
  {1956})}\BibitemShut {NoStop}%
\bibitem [{\citenamefont {Yosida}(1957)}]{Yosida1957MagneticAlloys}%
  \BibitemOpen
  \bibfield  {author} {\bibinfo {author} {\bibfnamefont {K.}~\bibnamefont
  {Yosida}},\ }\href@noop {} {\bibfield  {journal} {\bibinfo  {journal} {Phys.
  Rev.}\ }\textbf {\bibinfo {volume} {106}},\ \bibinfo {pages} {893} (\bibinfo
  {year} {1957})}\BibitemShut {NoStop}%
\bibitem [{\citenamefont {Chesi}\ and\ \citenamefont
  {Loss}(2010)}]{Chesi2010RKKYCouplings}%
  \BibitemOpen
  \bibfield  {author} {\bibinfo {author} {\bibfnamefont {S.}~\bibnamefont
  {Chesi}}\ and\ \bibinfo {author} {\bibfnamefont {D.}~\bibnamefont {Loss}},\
  }\href {https://doi.org/10.1103/PhysRevB.82.165303} {\bibfield  {journal}
  {\bibinfo  {journal} {Phys. Rev. B}\ }\textbf {\bibinfo {volume} {82}},\
  \bibinfo {pages} {165303} (\bibinfo {year} {2010})}\BibitemShut {NoStop}%
\bibitem [{\citenamefont {Chen}\ \emph {et~al.}(2021)\citenamefont {Chen},
  \citenamefont {Jiang}, \citenamefont {Shao}, \citenamefont {Holtz},
  \citenamefont {Odstrčil}, \citenamefont {Guizar-Sicairos}, \citenamefont
  {Hanke}, \citenamefont {Ganschow}, \citenamefont {Schlom},\ and\
  \citenamefont {Muller}}]{Chen2021}%
  \BibitemOpen
  \bibfield  {author} {\bibinfo {author} {\bibfnamefont {Z.}~\bibnamefont
  {Chen}}, \bibinfo {author} {\bibfnamefont {Y.}~\bibnamefont {Jiang}},
  \bibinfo {author} {\bibfnamefont {Y.-T.}\ \bibnamefont {Shao}}, \bibinfo
  {author} {\bibfnamefont {M.~E.}\ \bibnamefont {Holtz}}, \bibinfo {author}
  {\bibfnamefont {M.}~\bibnamefont {Odstrčil}}, \bibinfo {author}
  {\bibfnamefont {M.}~\bibnamefont {Guizar-Sicairos}}, \bibinfo {author}
  {\bibfnamefont {I.}~\bibnamefont {Hanke}}, \bibinfo {author} {\bibfnamefont
  {S.}~\bibnamefont {Ganschow}}, \bibinfo {author} {\bibfnamefont {D.~G.}\
  \bibnamefont {Schlom}},\ and\ \bibinfo {author} {\bibfnamefont {D.~A.}\
  \bibnamefont {Muller}},\ }\href@noop {} {\bibfield  {journal} {\bibinfo
  {journal} {Science}\ }\textbf {\bibinfo {volume} {372}},\ \bibinfo {pages}
  {826} (\bibinfo {year} {2021})}\BibitemShut {NoStop}%
\bibitem [{\citenamefont
  {Dresselhaus}(1955)}]{Dresselhaus1955Spin-OrbitStructures}%
  \BibitemOpen
  \bibfield  {author} {\bibinfo {author} {\bibfnamefont {G.}~\bibnamefont
  {Dresselhaus}},\ }\href@noop {} {\bibfield  {journal} {\bibinfo  {journal}
  {Phys. Rev.}\ }\textbf {\bibinfo {volume} {199}},\ \bibinfo {pages} {556}
  (\bibinfo {year} {1955})}\BibitemShut {NoStop}%
\bibitem [{\citenamefont {Bychkov}\ and\ \citenamefont
  {Rashba}(1984)}]{Bychkov1984OscillatoryLayers}%
  \BibitemOpen
  \bibfield  {author} {\bibinfo {author} {\bibfnamefont {Y.~A.}\ \bibnamefont
  {Bychkov}}\ and\ \bibinfo {author} {\bibfnamefont {E.~I.}\ \bibnamefont
  {Rashba}},\ }\href@noop {} {\bibfield  {journal} {\bibinfo  {journal} {J.
  Phys. C: Solid State Phys.}\ }\textbf {\bibinfo {volume} {17}},\ \bibinfo
  {pages} {6039} (\bibinfo {year} {1984})}\BibitemShut {NoStop}%
\bibitem [{sup()}]{supp}%
  \BibitemOpen
  \href@noop {} {}\bibinfo {note} {{See Supplemental Material for detailed
  calculations, which includes Refs. [], at
  http://link.aps.org/supplemental/...}}\BibitemShut {Stop}%
\bibitem [{\citenamefont {Berman}\ and\ \citenamefont
  {Flatt{\'{e}}}(2010)}]{Berman2010Electron-beamWells}%
  \BibitemOpen
  \bibfield  {author} {\bibinfo {author} {\bibfnamefont {D.~H.}\ \bibnamefont
  {Berman}}\ and\ \bibinfo {author} {\bibfnamefont {M.~E.}\ \bibnamefont
  {Flatt{\'{e}}}},\ }\href@noop {} {\bibfield  {journal} {\bibinfo  {journal}
  {Phys. Rev. Lett}\ }\textbf {\bibinfo {volume} {105}},\ \bibinfo {pages}
  {157202} (\bibinfo {year} {2010})}\BibitemShut {NoStop}%
\bibitem [{\citenamefont {Fiete}\ and\ \citenamefont
  {Heller}(2003)}]{Fiete2003Colloquium:Mirages}%
  \BibitemOpen
  \bibfield  {author} {\bibinfo {author} {\bibfnamefont {G.~A.}\ \bibnamefont
  {Fiete}}\ and\ \bibinfo {author} {\bibfnamefont {E.~J.}\ \bibnamefont
  {Heller}},\ }\href {http://www.almaden.ibm.com/} {\bibfield  {journal}
  {\bibinfo  {journal} {Rev. Mod. Phys.}\ }\textbf {\bibinfo {volume} {75}},\
  \bibinfo {pages} {933} (\bibinfo {year} {2003})}\BibitemShut {NoStop}%
\bibitem [{\citenamefont {Bouaziz}\ \emph {et~al.}(2018)\citenamefont
  {Bouaziz}, \citenamefont {dos Santos~Dias}, \citenamefont {Guimar{\~{a}}es},
  \citenamefont {Bl{\"{u}}gel},\ and\ \citenamefont
  {Lounis}}]{Bouaziz2018Impurity-inducedGas}%
  \BibitemOpen
  \bibfield  {author} {\bibinfo {author} {\bibfnamefont {J.}~\bibnamefont
  {Bouaziz}}, \bibinfo {author} {\bibfnamefont {M.}~\bibnamefont {dos
  Santos~Dias}}, \bibinfo {author} {\bibfnamefont {F.~S.~M.}\ \bibnamefont
  {Guimar{\~{a}}es}}, \bibinfo {author} {\bibfnamefont {S.}~\bibnamefont
  {Bl{\"{u}}gel}},\ and\ \bibinfo {author} {\bibfnamefont {S.}~\bibnamefont
  {Lounis}},\ }\href@noop {} {\bibfield  {journal} {\bibinfo  {journal} {Phys.
  Rev. B}\ }\textbf {\bibinfo {volume} {98}},\ \bibinfo {pages} {125420}
  (\bibinfo {year} {2018})}\BibitemShut {NoStop}%
\bibitem [{\citenamefont {Bernevig}\ \emph {et~al.}(2006)\citenamefont
  {Bernevig}, \citenamefont {Orenstein},\ and\ \citenamefont
  {Zhang}}]{Bernevig2006ExactSystem}%
  \BibitemOpen
  \bibfield  {author} {\bibinfo {author} {\bibfnamefont {B.~A.}\ \bibnamefont
  {Bernevig}}, \bibinfo {author} {\bibfnamefont {J.}~\bibnamefont
  {Orenstein}},\ and\ \bibinfo {author} {\bibfnamefont {S.-C.}\ \bibnamefont
  {Zhang}},\ }\href@noop {} {\bibfield  {journal} {\bibinfo  {journal} {Phys.
  Rev. Lett.}\ }\textbf {\bibinfo {volume} {97}},\ \bibinfo {pages} {236601}
  (\bibinfo {year} {2006})}\BibitemShut {NoStop}%
\bibitem [{\citenamefont {Jackson}(1999)}]{Jackson}%
  \BibitemOpen
  \bibfield  {author} {\bibinfo {author} {\bibfnamefont {J.~D.}\ \bibnamefont
  {Jackson}},\ }\href@noop {} {\emph {\bibinfo {title} {{Classical
  electrodynamics}}}},\ \bibinfo {edition} {3rd}\ ed.\ (\bibinfo  {publisher}
  {Wiley},\ \bibinfo {year} {1999})\BibitemShut {NoStop}%
\bibitem [{\citenamefont {Grundler}(2000)}]{Grundler2000LargeLayers}%
  \BibitemOpen
  \bibfield  {author} {\bibinfo {author} {\bibfnamefont {D.}~\bibnamefont
  {Grundler}},\ }\href@noop {} {\bibfield  {journal} {\bibinfo  {journal}
  {Phys. Rev. Lett}\ }\textbf {\bibinfo {volume} {84}},\ \bibinfo {pages}
  {6074} (\bibinfo {year} {2000})}\BibitemShut {NoStop}%
\bibitem [{\citenamefont {Yu}\ \emph {et~al.}(2016)\citenamefont {Yu},
  \citenamefont {Zeng}, \citenamefont {Cheng}, \citenamefont {Chen},
  \citenamefont {Liu}, \citenamefont {Lai}, \citenamefont {Zheng},\ and\
  \citenamefont {Ren}}]{Yu2016TuningWells}%
  \BibitemOpen
  \bibfield  {author} {\bibinfo {author} {\bibfnamefont {J.}~\bibnamefont
  {Yu}}, \bibinfo {author} {\bibfnamefont {X.}~\bibnamefont {Zeng}}, \bibinfo
  {author} {\bibfnamefont {S.}~\bibnamefont {Cheng}}, \bibinfo {author}
  {\bibfnamefont {Y.}~\bibnamefont {Chen}}, \bibinfo {author} {\bibfnamefont
  {Y.}~\bibnamefont {Liu}}, \bibinfo {author} {\bibfnamefont {Y.}~\bibnamefont
  {Lai}}, \bibinfo {author} {\bibfnamefont {Q.}~\bibnamefont {Zheng}},\ and\
  \bibinfo {author} {\bibfnamefont {J.}~\bibnamefont {Ren}},\ }\href@noop {}
  {\bibfield  {journal} {\bibinfo  {journal} {Nanoscale Res. Lett.}\ }\textbf
  {\bibinfo {volume} {11}},\ \bibinfo {pages} {477} (\bibinfo {year}
  {2016})}\BibitemShut {NoStop}%
\bibitem [{\citenamefont {Nitta}\ \emph {et~al.}(1997)\citenamefont {Nitta},
  \citenamefont {Akazaki}, \citenamefont {Takayanagi},\ and\ \citenamefont
  {Enoki}}]{Nitta1997GateHeterostructure}%
  \BibitemOpen
  \bibfield  {author} {\bibinfo {author} {\bibfnamefont {J.}~\bibnamefont
  {Nitta}}, \bibinfo {author} {\bibfnamefont {T.}~\bibnamefont {Akazaki}},
  \bibinfo {author} {\bibfnamefont {H.}~\bibnamefont {Takayanagi}},\ and\
  \bibinfo {author} {\bibfnamefont {T.}~\bibnamefont {Enoki}},\ }\href@noop {}
  {\bibfield  {journal} {\bibinfo  {journal} {Phys. Rev. Lett.}\ }\textbf
  {\bibinfo {volume} {78}},\ \bibinfo {pages} {1335} (\bibinfo {year}
  {1997})}\BibitemShut {NoStop}%
\bibitem [{\citenamefont {Bahrami}\ \emph {et~al.}(2021)\citenamefont
  {Bahrami}, \citenamefont {Houshang}, \citenamefont {Ju}, \citenamefont {Bie},
  \citenamefont {Shang}, \citenamefont {Jamdagni}, \citenamefont {Pandey},\
  and\ \citenamefont {Tankeshwar}}]{Bahrami2021}%
  \BibitemOpen
  \bibfield  {author} {\bibinfo {author} {\bibfnamefont {A.}~\bibnamefont
  {Bahrami}}, \bibinfo {author} {\bibfnamefont {A.}~\bibnamefont {Houshang}},
  \bibinfo {author} {\bibfnamefont {L.}~\bibnamefont {Ju}}, \bibinfo {author}
  {\bibfnamefont {M.}~\bibnamefont {Bie}}, \bibinfo {author} {\bibfnamefont
  {J.}~\bibnamefont {Shang}}, \bibinfo {author} {\bibfnamefont
  {P.}~\bibnamefont {Jamdagni}}, \bibinfo {author} {\bibfnamefont
  {R.}~\bibnamefont {Pandey}},\ and\ \bibinfo {author} {\bibfnamefont
  {K.}~\bibnamefont {Tankeshwar}},\ }\href
  {https://doi.org/10.1088/1361-6528/AC2D46} {\bibfield  {journal} {\bibinfo
  {journal} {Nanotechnology}\ }\textbf {\bibinfo {volume} {33}},\ \bibinfo
  {pages} {025703} (\bibinfo {year} {2021})}\BibitemShut {NoStop}%
\bibitem [{\citenamefont {van Bree}\ \emph {et~al.}(2014)\citenamefont {van
  Bree}, \citenamefont {\hbox{Yu} Silov}, \citenamefont {Koenraad},\ and\
  \citenamefont {Flatt{\'{e}}}}]{VanBree2014Spin-orbit-inducedNanostructure}%
  \BibitemOpen
  \bibfield  {author} {\bibinfo {author} {\bibfnamefont {J.}~\bibnamefont {van
  Bree}}, \bibinfo {author} {\bibfnamefont {A.}~\bibnamefont {\hbox{Yu}
  Silov}}, \bibinfo {author} {\bibfnamefont {P.~M.}\ \bibnamefont {Koenraad}},\
  and\ \bibinfo {author} {\bibfnamefont {M.~E.}\ \bibnamefont {Flatt{\'{e}}}},\
  }\href@noop {} {\bibfield  {journal} {\bibinfo  {journal} {Phys. Rev. Lett}\
  }\textbf {\bibinfo {volume} {112}},\ \bibinfo {pages} {187201} (\bibinfo
  {year} {2014})}\BibitemShut {NoStop}%
\end{thebibliography}%
\bibliographystyle{apsrev4-2}

\end{document}


\title{Dissipationless Circulating Currents and  Fringe Magnetic Fields Near a Single Spin Embedded in a  Two-Dimensional Electron Gas \\ Supplemental Material}

\author{Adonai R. da Cruz}
\affiliation{Department of Applied Physics, Eindhoven University of Technology, Eindhoven 5612 AZ, The Netherlands}

\author{Michael E. Flatt\'{e}}
\affiliation{Department of Physics and Astronomy, Optical Science and Technology Center, University of Iowa, Iowa City, Iowa 52242, USA}
\affiliation{Department of Applied Physics, Eindhoven University of Technology, Eindhoven 5612 AZ, The Netherlands}

\setcounter{equation}{0}
\setcounter{figure}{0}
\setcounter{table}{0}
\setcounter{page}{1}
\makeatletter
\renewcommand{\theequation}{S\arabic{equation}}
\renewcommand{\thefigure}{S\arabic{figure}}
\renewcommand{\bibnumfmt}[1]{[S#1]}
\renewcommand{\citenumfont}[1]{S#1}


\maketitle

\section{Derivation of the current operator}

To derive a current operator which is consistent with the continuity
equation by construction, we start by taking the time derivative of
the density operator $\hat{n}=\left|\psi\right|^{2}$ and using the
time-dependent Schr\"{o}dinger equation to obtain,

\begin{align}
\frac{\partial\hat{n}}{\partial t} & =-\frac{1}{i\hbar}\left(\mathbf{\hat{H}}^{\dagger}\psi^{\dagger}\right)\psi+\psi^{\dagger}\frac{1}{i\hbar}\left(\hat{\mathbf{H}}\psi\right).\\
\nonumber 
\end{align}
After substituting the Hamiltonian of Eq.~(1) of the main text and
using the real-space representation of the momentum operator, $\hat{p}=-i\hbar\nabla$,
the above equation can be written as

\begin{align}
\frac{\partial\hat{n}}{\partial t} & =\left[-\frac{\hbar}{2im}\psi^{\dagger}\left(\nabla^{2}\psi\right)+\frac{\hbar}{2im}\left(\nabla^{2}\psi^{\dagger}\right)\psi\right]\nonumber \\
 & +\frac{1}{\hbar}\left[\alpha\sigma_{x}\left(\partial_{y}\psi^{\dagger}\right)-\alpha\sigma_{y}\left(\partial_{x}\psi^{\dagger}\right)+\beta\sigma_{x}\left(\partial_{x}\psi^{\dagger}\right)-\beta\sigma_{y}\left(\partial_{y}\psi^{\dagger}\right)\right]\psi\nonumber \\
 & +\frac{1}{\hbar}\psi^{\dagger}\left[\alpha\sigma_{x}\left(\partial_{y}\psi\right)-\alpha\sigma_{y}\left(\partial_{x}\psi\right)+\beta\sigma_{x}\left(\partial_{x}\psi\right)-\beta\sigma_{y}\left(\partial_{y}\psi\right)\right],
\end{align}
and the spin-orbit terms can be grouped together by defining a spin-dependent
vector potential, given in cartesian coordinates, by

\begin{equation}
\mathbf{A}_{so}=\frac{m^{*}}{e\hbar}\left(\beta\sigma_{x}-\alpha\sigma_{y},\alpha\sigma_{x}-\beta\sigma_{y}\right),
\end{equation}
such that, 

\begin{equation}
\frac{\partial\hat{n}}{\partial t}=-\frac{\hbar}{2im}\left[\psi^{\dagger}\left(\nabla^{2}\psi\right)-\left(\nabla^{2}\psi^{\dagger}\right)\psi\right]+\frac{e}{m^{*}}\left[A_{so}^{x}\psi^{\dagger}\left(\partial_{x}\psi\right)+A_{so}^{y}\psi^{\dagger}\left(\partial_{y}\psi\right)+A_{so}^{x}\left(\partial_{x}\psi^{\dagger}\right)\psi+A_{so}^{y}\left(\partial_{y}\psi^{\dagger}\right)\psi\right].
\end{equation}
 Using vector relations, the time derivative of the density operator
can be recast

\[
\frac{\partial\hat{n}}{\partial t}=-\frac{\hbar}{2im^{*}}\nabla\cdot\left(\psi^{\dagger}\nabla\psi-\left(\nabla\psi^{\dagger}\right)\psi\right)+\frac{e}{m^{*}}\nabla\cdot\left(\psi^{\dagger}\mathbf{A}_{so}\psi\right).
\]
The current operator is obtained by comparing the above expression with the continuity equation, $\partial_{t}\hat{n}=-\nabla\cdot\hat{j}$,

\begin{equation}
\hat{j}=\frac{\hbar}{2im^{*}}\left(\psi^{\dagger}\nabla\psi-\left(\nabla\psi^{\dagger}\right)\psi\right)-\frac{e}{m^{*}}\left(\psi^{\dagger}\mathbf{A}_{so}\psi\right).
\end{equation}
To evaluate the charge current operator using retarded Green's functions, we
use the spectral representation,

\[
G\left(\mathbf{r},\mathbf{r}^{\prime};E+i0^{+}\right)=\sum_{n}\frac{\psi_{n}^{\dagger}\left(\mathbf{r}\right)\psi_{n}\left(\mathbf{r}^{\prime}\right)}{E-E_{n}+i0^{+}}
\]
and integrate in energy over all occupied states to obtain

\begin{equation}
\mathbf{j}\left(\mathbf{r}\right)=-\frac{e\hbar}{2\pi m^{*}}\text{Im}\int_{E_{so}}^{E_{f}}\mathrm{Tr}\left\{ i\lim_{\mathbf{r}^{\prime}\to\mathbf{r}}\left(\nabla-\nabla^{\prime}\right)G\left(\mathbf{r},\mathbf{r}^{\prime};E\right)-\frac{2e}{\hbar}\lim_{\mathbf{r}^{\prime}\to\mathbf{r}}\mathbf{A}_{so}G\left(\mathbf{r},\mathbf{r}^{\prime};E\right)\right\} dE\label{eq:current}
\end{equation}
where $E_{so}$ is the bottom of the spin-split bands. Within the
$s$-wave approximation for the T-matrix, the magnetic potential for
a defect located at the origin is included in the Green's function
formalism by the Dyson equation,

\[
\mathbf{G}\left(\mathbf{r},\mathbf{r}^{\prime};E\right)=\mathbf{G}^{0}\left(\mathbf{r},0;E\right)\mathbf{T}\mathbf{G}^{0}\left(0,\mathbf{r}^{\prime};E\right),
\]
with

\[
\mathbf{T}=\left(\begin{array}{cc}
t_{\uparrow} & 0\\
0 & t_{\downarrow}
\end{array}\right).
\]
From the rotation symmetries of the homogeneous Green's
functions, we have, for spin diagonal terms,

\begin{equation}
G_{\sigma,\sigma}^{0}\left(\mathbf{r},0;E\right)=G_{\sigma,\sigma}^{0}\left(-\mathbf{r},0;E\right)=G_{\sigma,\sigma}^{0}\left(0,\mathbf{r};E\right),
\end{equation}
while the off-diagonal terms,

\begin{equation}
G_{\sigma,\sigma^{\prime}}^{0}\left(\mathbf{r},0;E\right)=-G_{\sigma,\sigma^{\prime}}^{0}\left(-\mathbf{r},0;E\right)=-G_{\sigma,\sigma^{\prime}}^{0}\left(0,\mathbf{r};E\right).
\end{equation}
Using these relations, the defect Green's functions are obtained:

\begin{align}
G_{\uparrow\uparrow}\left(\mathbf{r}\right) & =t_{\uparrow}\left[G_{\uparrow\uparrow}^{0}\left(\mathbf{r}\right)\right]^{2}-t_{\downarrow}G_{\uparrow\downarrow}^{0}\left(\mathbf{r}\right)G_{\downarrow\uparrow}^{0}\left(\mathbf{r}\right)\\
G_{\uparrow\downarrow}\left(\mathbf{r}\right) & =-\left(t_{\uparrow}-t_{\downarrow}\right)G_{\uparrow\uparrow}^{0}\left(\mathbf{r}\right)G_{\uparrow\downarrow}^{0}\left(\mathbf{r}\right)\\
G_{\downarrow\uparrow}\left(\mathbf{r}\right) & =\left(t_{\uparrow}-t_{\downarrow}\right)G_{\uparrow\uparrow}^{0}\left(\mathbf{r}\right)G_{\downarrow\uparrow}^{0}\left(\mathbf{r}\right)\\
G_{\downarrow\downarrow}\left(\mathbf{r}\right) & =t_{\downarrow}\left[G_{\downarrow\downarrow}^{0}\left(\mathbf{r}\right)\right]^{2}-t_{\uparrow}G_{\downarrow\uparrow}^{0}\left(\mathbf{r}\right)G_{\uparrow\downarrow}^{0}\left(\mathbf{r}\right)
\end{align}
The expectation value of the charge current associated with a spin-polarized
defect is calculated by substituting the above expressions for the
defect Green's functions in Eq.~(\ref{eq:current}),

\begin{equation}
\mathbf{j}\left(\mathbf{r}\right)=-\frac{e\hbar}{2\pi m^{*}}\text{Im}\int_{E_{so}}^{E_{f}}\left(t_{\uparrow}-t_{\downarrow}\right)\left\{ i\left[\left(\nabla G_{\uparrow\downarrow}^{0}\right)G_{\downarrow\uparrow}^{0}-\left(\nabla G_{\downarrow\uparrow}^{0}\right)G_{\uparrow\downarrow}^{0}\right]-\frac{2e}{\hbar}\left(\mathbf{A}_{so}^{\uparrow\downarrow}G_{\downarrow\uparrow}^{0}-\mathbf{A}_{so}^{\downarrow\uparrow}G_{\uparrow\downarrow}^{0}\right)G_{\uparrow\uparrow}^{0}\right\} dE
\end{equation}

\section{Gradients of the homogeneous Greens' functions}

The current operator derived in this work depends on the gradient
of the 2DEG Green's functions. To evaluate those terms in cylindrical
coordinates, we will employ $\nabla=\partial_{r}+(1/r)\partial_{\theta}$
to find,

\begin{align*}
\nabla G_{\uparrow\downarrow}^{0}\cdot\hat{r} & =\frac{\partial}{\partial r}G_{\uparrow\downarrow}^{0}\left(\mathbf{r};E\right)\\
 & =\frac{im}{2\pi^{2}\hbar^{2}}\int_{\theta-\pi/2}^{\theta+\pi/2}d\theta_{k}\frac{\left(U_{x}+iU_{y}\right)}{Q\sqrt{f_{\tau}}}\left[k_{+}^{2}I_{1}\left(k_{+}\rho\right)-k_{-}^{2}I_{1}\left(k_{-}\rho\right)\right]\text{cos}\left(\theta_{k}-\theta\right)
\end{align*}

\begin{align*}
\nabla G_{\uparrow\downarrow}^{0}\cdot\hat{\theta} & =\frac{1}{r}\frac{\partial}{\partial\theta}G_{\uparrow\downarrow}^{0}\left(\mathbf{r};E\right)\\
 & =\frac{im}{2\pi^{2}\hbar^{2}}\int_{\theta-\pi/2}^{\theta+\pi/2}d\theta_{k}\frac{\left(U_{x}+iU_{y}\right)}{Q\sqrt{f_{\tau}}}\left[k_{+}^{2}I_{1}\left(k_{+}\rho\right)-k_{-}^{2}I_{1}\left(k_{-}\rho\right)\right]\text{sin}\left(\theta_{k}-\theta\right)
\end{align*}

and

\begin{align*}
\nabla G_{\downarrow\uparrow}\cdot\hat{r} & =\frac{\partial}{\partial r}G_{\downarrow\uparrow}^{0}\left(\mathbf{r};E\right)\\
 & =\frac{im}{2\pi^{2}\hbar^{2}}\int_{\theta-\pi/2}^{\theta+\pi/2}d\theta_{k}\frac{\left(U_{x}-iU_{y}\right)}{Q\sqrt{f_{\tau}}}\left[k_{+}^{2}I_{1}\left(k_{+}\rho\right)-k_{-}^{2}I_{1}\left(k_{-}\rho\right)\right]\text{cos}\left(\theta_{k}-\theta\right)
\end{align*}

\begin{align*}
\nabla G_{\downarrow\uparrow}^{0}\cdot\hat{\theta} & =\frac{1}{r}\frac{\partial}{\partial\theta}G_{\downarrow\uparrow}^{0}\left(\mathbf{r};E\right)\\
 & =\frac{im}{2\pi^{2}\hbar^{2}}\int_{\theta-\pi/2}^{\theta+\pi/2}d\theta_{k}\frac{\left(U_{x}-iU_{y}\right)}{Q\sqrt{f_{\tau}}}\left[k_{+}^{2}I_{1}\left(k_{+}\rho\right)-k_{-}^{2}I_{1}\left(k_{-}\rho\right)\right]\text{sin}\left(\theta_{k}-\theta\right)
\end{align*}

\section{Effective spin-orbit vector potential in polar coordinates}

Since we have calculated the Green's functions in polar coordinates,
we also need to perform a coordinate transformation of the spin-orbit gauge. Using the unit vectors in cartesian coordinates, 
$\hat{r}=\left(\cos\theta,\sin\theta\right)$ and $\hat{\theta}=\left(-\sin\theta,\cos\theta\right)$
and the component of $\mathbf{A}$ in cartesian coordinates,

\[
A_{so}^{x}=\frac{\hbar k_{so}}{e}\left(\begin{array}{cc}
0 & e^{i\tau}\\
e^{-i\tau} & 0
\end{array}\right),\quad A_{so}^{y}=\frac{\hbar k_{so}}{e}\left(\begin{array}{cc}
0 & ie^{-i\tau}\\
-ie^{i\tau} & 0
\end{array}\right)
\]
Applying a transformation to polar coordinates defined by $A_{so}^{r}=A_{so}^{x}\cos\theta+A_{so}^{y}\sin\theta$
and $A_{so}^{\theta}=A_{so}^{y}\cos\theta-A_{so}^{x}\sin\theta$,
to the above expressions,

\[
A_{so}^{r}=\frac{\hbar k_{so}}{e}\left(\begin{array}{cc}
0 & e^{i\tau}\cos\theta+ie^{-i\tau}\sin\theta\\
e^{-i\tau}\cos\theta-ie^{i\tau}\sin\theta & 0
\end{array}\right),
\]
and

\[
A_{so}^{\theta}=\frac{\hbar k_{so}}{e}\left(\begin{array}{cc}
0 & ie^{-i\tau}\cos\theta-e^{i\tau}\sin\theta\\
-ie^{i\tau}\cos\theta-e^{-i\tau}\sin\theta & 0
\end{array}\right).
\]
Using Pauli matrices we can write these components in a more compact
form as shown in the main text,

\[
A_{so}^{r}=\frac{\hbar k_{so}}{e}\left[\sigma_{x}\cos\left(\theta-\tau\right)-\sigma_{y}\sin\left(\theta+\tau\right)\right],
\]

\[
A_{so}^{\theta}=\frac{\hbar k_{so}}{e}\left[\sigma_{x}\sin\left(\theta-\tau\right)-\sigma_{y}\cos\left(\theta+\tau\right)\right].
\]

\section{Current circulation symmetry}

Here we will show the circulation symmetry that exists for a SOC angle
$\tau$ around $\pi/4$ (corresponding to $\alpha=\beta$). The off-diagonal Green's functions are related by phases given by $U_{x}\pm iU_{y}$. Let us define two symmetric angles: $\tau^{+}=\pi/4+\delta$ and
$\tau^{-}=\pi/4-\delta$. We notice that,

\begin{align}
U_{x}^{\pm} & =\cos\left(\pi/4\pm\delta\right)\cos\left(\theta_{k}\right)+\sin\left(\pi/4\pm\text{}\delta\right)\sin\left(\theta_{k}\right),\\
U_{y}^{\pm} & =\sin\left(\pi/4\pm\delta\right)\cos\left(\theta_{k}\right)+\cos\left(\pi/4\pm\delta\right)\sin\left(\theta_{k}\right).
\end{align}
Using trigonometric identities it is straightforward to show that $U_{x}^{+}=U_{y}^{-}$
and $U_{y}^{+}=U_{x}^{-}$ . From Eq.~(2) in the main text
the diamagnetic contribution to the current is proportional
to the product between off-diagonal Green's functions. The relation between these
products evaluated at $\tau^{+}$ and $\tau^{-}$ is

\begin{align}
\left(\nabla G_{\uparrow\downarrow}^{0}G_{\downarrow\uparrow}^{0}\right)^{+} & =\left(U_{x}^{+}+iU_{y}^{+}\right)\left(U_{x}^{+}-iU_{y}^{+}\right)\nonumber \\
 & =\left(U_{y}^{-}+iU_{x}^{-}\right)\left(U_{y}^{-}-iU_{x}^{-}\right)\nonumber \\
 & =-\left(U_{x}^{-}-iU_{y}^{-}\right)\left(U_{x}^{-}+iU_{y}^{-}\right)\nonumber \\
 & =-\left(\nabla G_{\downarrow\uparrow}^{0}G_{\uparrow\downarrow}^{0}\right)^{-}
\end{align}
and $\left(\nabla G_{\downarrow\uparrow}^{0}G_{\uparrow\downarrow}^{0}\right)^{+}=-\left(\nabla G_{\uparrow\downarrow}^{0}G_{\downarrow\uparrow}^{0}\right)^{-}$.
To show that similar relations exist for the paramagnetic-like term
in Eq.~(2) in the main text, we evaluate here only the radial component
of the spin-dependent vector potential $\mathbf{A}_{so}$, the same
procedure equally applies to the angular component. Note that we
can write,

\begin{equation}
\mathbf{A}_{so}^{\uparrow\downarrow}\cdot\hat{r}=\frac{\hbar^{2}k_{so}}{m}\left(U_{x}+iU_{y}\right),\qquad\mathbf{A}_{so}^{\downarrow\uparrow}\cdot\hat{r}=\frac{\hbar^{2}k_{so}}{m}\left(U_{x}-iU_{y}\right).
\end{equation}
After performing the substitutions for $U_{x}$ and $U_{y}$
we find the relations $\left(\mathbf{A}_{so}^{\uparrow\downarrow}G_{\downarrow\uparrow}^{0}\right)^{+}=-\left(\mathbf{A}_{so}^{\uparrow\downarrow}G_{\downarrow\uparrow}^{0}\right)^{-}$
and $\left(\mathbf{A}_{so}^{\uparrow\downarrow}G_{\downarrow\uparrow}^{0}\right)^{+}=-\left(\mathbf{A}_{so}^{\uparrow\downarrow}G_{\downarrow\uparrow}^{0}\right)^{-}$.
We thus conclude that the currents at $\tau^{+}$ and $\tau^{-}$
have the same magnitude but circulate in opposite directions to each other.